\documentclass{aa}

\usepackage{graphicx}
\usepackage{txfonts}
\usepackage{skak}
\usepackage{multirow}
\usepackage{color}
\usepackage{paralist}
\usepackage{amsmath}
\usepackage{tablefootnote}
\usepackage{amssymb}
\usepackage{bm}
\usepackage{tabularx}
\usepackage{xcolor}
\usepackage[colorlinks=true,allcolors=blue]{hyperref}
\bibpunct{(}{)}{;}{a}{}{,}

\def\teff{$\textit{T}_{\rm{eff}}$}

\def\logg{\mbox{log~{\it g}}}
\def\vmicro{\mbox{$\xi_{\rm t}$}}

\begin{document} 
   \title{\king Playing CHESS with stars}
   \subtitle{II. Homogeneous analysis of high-resolution optical spectra for a sample of 135 giants in 14 bulge globular clusters\thanks{Based on data obtained from the ESO Science Archive Facility.}}
   \author{S.~\"{O}zdemir
          \inst{1}
          \and
          R.~Smiljanic\inst{1}          
          \and
          J.~E.~Mart\'{\i}nez Fern\'andez\inst{1}
          \and
          P.~R.~T.~Coelho\inst{2}
         \and
         F.~Primas\inst{3}
          \and
          E.~Valenti\inst{3}
     }

   \institute{Nicolaus Copernicus Astronomical Center, 
              Polish Academy of Sciences, ul. Bartycka 18, 00-716, Warsaw, Poland\\
              \email{sergenozdemir58@gmail.com; sergen@camk.edu.pl}
              \and 
              Universidade de S\~ao Paulo, IAG, Rua do Mat\~ao, 1226, 05508-090, S\~ao Paulo, SP, Brazil
              \and
              European Southern Observatory, 
              Karl-Schwarzschild Strasse 2, 85748 Garching bei M\"unchen, Germany        
             }
   \date{Received --; accepted --}

  \abstract
   {Bulge globular clusters are key tracers of the complex formation history of the Galactic bulge, as they retain information on its past star-formation episodes. Their chemical abundances serve as empirical “fossil records” of the early evolution of the bulge.}
   {We present a homogeneous analysis of archival high-resolution UVES spectra for 135 giants in 14 Galactic bulge globular clusters. We explore the results in the context of the phenomenon of multiple stellar populations and discuss what the abundances reveal about the chemical evolution of the bulge.}
   {We employed the CHEmical Survey analysis System (CHESS) pipeline to derive atmospheric parameters on an accurate scale defined by reference stars. Abundances for Li, O, Na, Mg, Al, Si, Ca, Sc, Y, Zr, Ba, Ce, Nd, and Eu were derived by spectrum synthesis. Deviations from local thermodynamic equilibrium (i.e., non-LTE or NLTE) were taken into account for Na, Mg, Al, Si, Ca, Y, Ba, and Eu.}
   {We serendipitously discovered a Li-rich giant with A(Li) = 1.68 (after 3D NLTE corrections) in NGC 6723, which was hidden in the archival data. Our sample includes both first- and second-generation stars, as indicated by the observed spread in Na abundances. Compared to halo clusters, the bulge clusters show a lower [$\alpha$/Fe] plateau at low metallicity and a more gradual decline in [$\alpha$/Fe] toward higher metallicities. The Sc abundances are below solar and lower than in disk stars of the same metallicity. This difference could be important for solving open questions about the nucleosynthesis of this element. None of the clusters exhibit any significant spreads in heavy elements. Most clusters exhibit [Ba/Eu] $\sim$ 0.0, indicating a contribution from both s- and r-processes. Clusters with [Fe/H] $<$ $-$0.5 show negative [Ce/Nd] ratios, supporting the idea that these elements were predominantly produced by the r-process in these clusters.}
   {}

   \keywords{Stars: abundances --
   Galaxy: bulge -- (Galaxy:) globular clusters: general
            }

    \titlerunning{Homogeneous analysis of bulge globular clusters}
    \authorrunning{S. \"Ozdemir et al.}
    \maketitle
    \nolinenumbers

\section{Introduction}\label{sec:intro}

The Galactic bulge was recognized as a distinct structural component of the Milky Way eight decades ago \citep{Baade1946, Stebbins1947}. Since then, the processes driving its formation, evolution, and chemical enrichment have been a topic of growing interest \citep[see reviews by][]{Barbuy_review, Zoccali2019rev}. Processes such as gas accretion, early mergers, and inward migration of thick-disk gas clumps are likely to have played a significant role in its formation, operating at different timescales. Nevertheless, there is still no fully self-consistent model for the Galactic bulge formation \citep{Barbuy_review}. The inner regions of the Milky Way are difficult to investigate as a result of severe interstellar extinction and stellar crowding. Recent progress in understanding the bulge is being driven by modern large photometric \citep[VISTA Variables in the Vía Láctea, VVV and VVV eXtended, VVVX;][respectively]{VVV, VVVX}, astrometric \citep[\textit{Gaia};][]{gaia16, gaiadr3}, and spectroscopic \citep[Apache Point Observatory Galactic Evolution Experiment, APOGEE;][]{apogee} surveys, some of them exploring infrared wavelength ranges where the effects of extinction are less pronounced. The new instrument MOONS \citep[Multi-Object Optical and Near-infrared Spectrograph,][]{MOONS} will contribute to that in the near future.

The bulge stellar population seems to consist of two main components, a metal-poor one ([Fe/H]\footnote{$\mathrm{[X/Y]=\log(N_X/N_Y)_{*}-\log(N_X/N_Y)_{\odot}}$} $\sim$ $-$0.5) distributed as a spheroidal and a metal-rich one ([Fe/H] $\sim$ +0.3) that dominates in the mid-plane and is responsible for the boxy/peanut shaped bar \citep{Hill2011, Rojas-Arriagada2017, Johnson2022}. \citet{MillerJoyce2025} compared APOGEE data with a two infall Galactic chemical evolution model implemented in the OMEGA++ code and found  evidence that the bulge formed as a combination of a classical collapse followed by a later star formation episode that produced a younger stellar component. Similarly, the chemical evolution modeling of \cite{Molero2024} could reproduce the metallicity distribution of bulge stars with two in situ bursts of star formation. On top of that, there is an extended distribution of stars towards the very metal-poor regime ([Fe/H] $<$ $-$2.0) likely dominated by a contribution from accreted galaxies \citep{ArdernArentsen2024}.

Studies of bulge globular clusters (GCs) have played a key role in improving our understanding of the formation scenarios and chemodynamical properties of this Galactic component. For GCs, distances and ages can be better determined than for field stars. Early discussions of the connection between inner Galaxy GCs and the bulge focused on their metallicities and radial velocities \citep{Minniti1995, Cote1999}. Among these early studies, it is worth mentioning \cite{Ortolani1995}, who demonstrated that bulge GCs are old and used this to infer that the bulge had been part of the halo formation process and that it had undergone rapid chemical enrichment. Since then, numerous studies have focused on examining the properties of these clusters. \cite{Valenti2007, Valenti2010} presented a homogeneous photometric survey to derive magnitudes, reddening, distances, and photometric metallicities for almost 80\% of bulge clusters\footnote{\url{http://www.bo.astro.it/~GC/ir_archive/}}. Similarly, \cite{capos_bulge} presented a catalog of homogeneous chemical abundances derived from high-resolution spectra of stars in 18 bulge GCs. 

There are several open issues concerning the origin and properties of the bulge clusters. The nature of several recently discovered candidate GCs still needs to be confirmed, while ages and spectroscopic metallicities are also missing in many cases \citep{Bica24}. In addition, there is the problem of identifying the in situ and ex situ GCs. Interactions with the harsh and crowded environment of the Galactic center and the general orbital evolution of GCs complicate such efforts \citep{Ishchenko2023, Ishchenko2024, Pagnini2023}. Recently, \cite{Origlia2025} showed that Terzan 5 displays evolutionary, kinematic, and chemical properties that support the hypothesis of its in situ formation. Using the chemical abundances of V and Zn as markers, \cite{Bartolomei2026} favoured an in situ origin for NGC 6553, one of the most metal-rich GCs in the bulge. However, identifying the origin of a cluster is not an easy task, since about 30\% of the bulge clusters could, in fact, be an inner halo or thick disk clusters passing through the bulge location \citep{PerezVillegas2020}.

The recent bulge GC census by \cite{Bica24} listed 83 GCs possibly associated with the bulge. That same study also confirmed that the bulge GCs exhibit a bimodal metallicity distribution. This suggests that we are tracing multiple star formation episodes. In particular, the lack of an age-metallicity relationship for old clusters supports an situ origin for these systems \citep{Bica24}. A similar bimodal metallicity distribution is also observed in bulge field stars again suggesting different star formation episodes \citep{Lian2020, Molero2024}.

The presence of multiple populations (MPs) and variations in the chemical abundances of light elements (e.g., C, N, O, and Na) is a defining characteristic observed in Milky Way GCs \citep[see reviews by][and references therein]{Gratton2012, Bastian2018, Milone2022}. Globular clusters mainly host two distinct populations: stars that have abundances similar to those of halo stars of the same metallicity (so-called first-generation stars) and stars that have enhanced abundances of N and Na and depleted abundances of O and C (second-generation stars). These chemical differences likely come from the evolution of stars from the first generation, but the nature of these polluters has not been determined thus far \citep[e.g.][and references therein]{Bastian2018}. Multiple populations can also be seen as distinct photometric sequences that are directly linked to the spectroscopic observations \citep{Dondoglio2025}. In their work, \citet{Dondoglio2025} also reported that more massive GCs seem to have higher iron and barium spreads. The need for homogeneous spectroscopic data was very clear in their work, being necessary for a robust comparison of so many clusters.

High-precision, homogeneous analyses of large GC samples are essential to improve our understanding of the GC metallicity distribution, as well as the origins of the clusters and the origins of their multiple stellar populations. There are several recent examples of such studies \citep{Schiavon2024, capos_bulge, Carretta2026b, Bartolomei2026} and our current analysis is part of an ongoing effort to further progress in this direction. This work is the first part of a homogeneous analysis of a large sample of high-resolution archival spectra of stars in GCs. Here, we focus on bulge GCs and we will present the results for the remaining sample in forthcoming papers. Our analysis uses a new pipeline, the CHEmical Survey analysis System \citep[CHESS,][]{CHESSI}, designed to derive homogeneous chemical abundances from archival spectra. With CHESS, we analyzed FGK-type stars observed with the Ultraviolet and Visual Echelle Spectrograph \citep[UVES,][]{UVES} on the Unit Telescope 2 of the Very Large Telescope (VLT), Paranal Observatory, Chile. This paper is organized as follows. Section ~\ref{sec:data} summarizes the data selection, sample stars, and their membership in bulge GCs. The analysis methodology is described in Section ~\ref{sec:method}. We discuss our results in Section ~\ref{sec:discussion} and summarize our conclusions in Section ~\ref{sec:summary}.                           

\section{Data and sample}\label{sec:data}

\begin{figure}
    \centering
    \includegraphics[width=\linewidth]{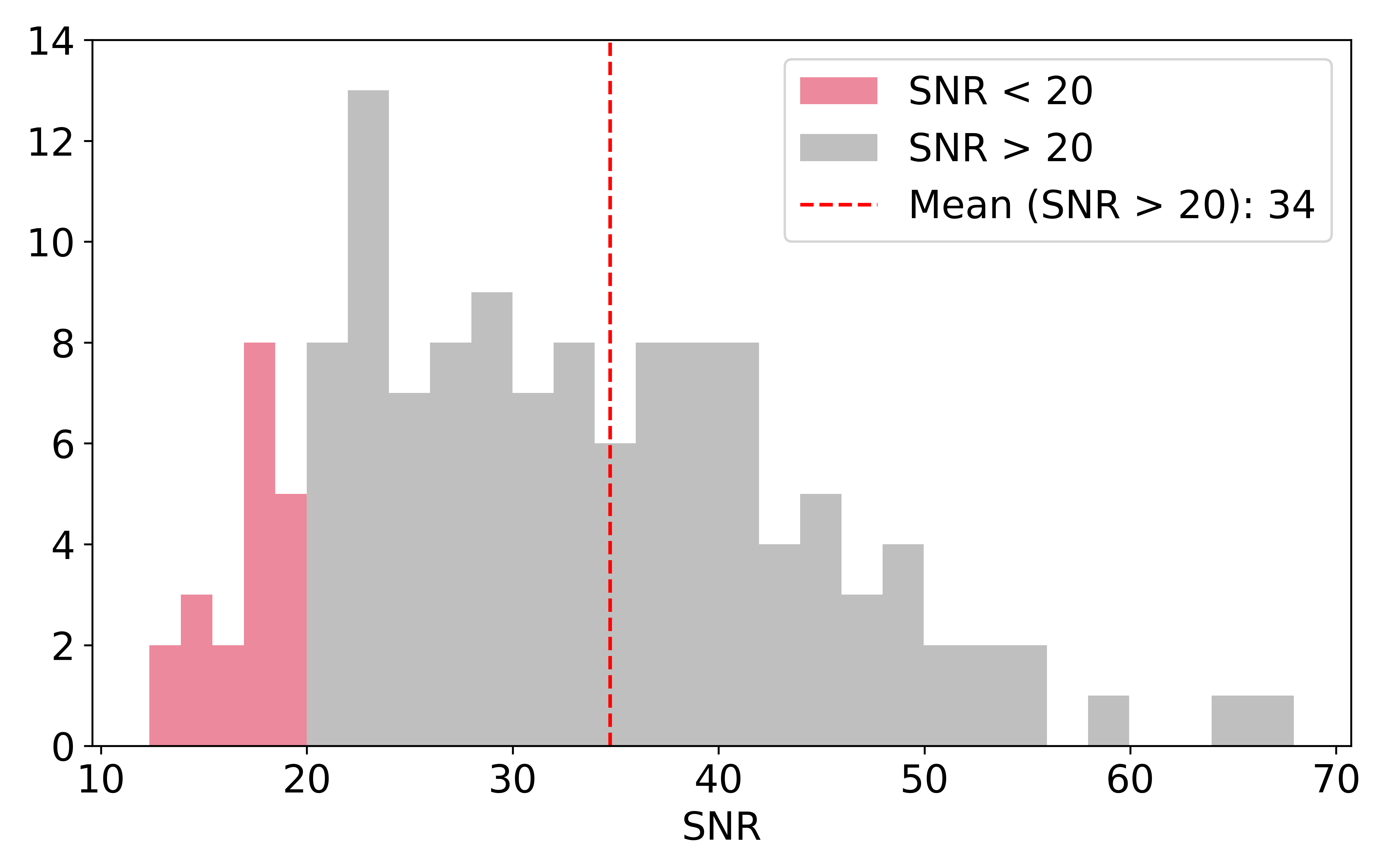}
    \caption{Histogram of S/N values for the sample of bulge globular cluster stars. We derived abundances for spectra S/N $>$ 20. Only atmospheric parameters were derived for the spectra with lower S/N (marked in red).}
    \label{fig:SNR}
\end{figure}

\begin{table*}[t]
\caption{Globular clusters studied in this work.}
\label{tab:clusters_summary}
$$
\begin{array}{lcccrrrcccc}
    \hline \hline
Cluster & RA^{\rm a} & DEC^{\rm a} & RV^{\rm b} & RV^{\rm a}  & RV^{\rm c} & RV^{\rm d} & Age^{\rm e} (Gyr) & [Fe/H]^{\rm a} & [Fe/H]^{\rm c} & [Fe/H]^{\rm d}\\
    \hline 
\rm HP\ 1     & 17\ 31\ 05.20 & -29\ 58\ 54.0 & 39.76  & 45.8   & 40.1   & 39.76   & 12.8\pm\ 0.9         & -1.00 & -1.23 & -1.21 \\
\rm NGC\ 6093 & 16\ 17\ 02.41 & -22\ 58\ 33.9 & 10.93  & 8.1    & --     & 10.93   & --                   & -1.75 & --    & -1.61 \\
\rm NGC\ 6171 & 16\ 32\ 31.86 & -13\ 03\ 13.6 & -34.71 & -34.1  & --     & -34.71  & --                   & -1.02 & --    & -1.02 \\
\rm NGC\ 6266 & 17\ 01\ 12.80 & -30\ 06\ 49.4 & -73.98  & -70.1  & --     & --      & 11.6\pm\ 0.6         & -1.18 & --    & --    \\
\rm NGC\ 6304 & 17\ 14\ 32.25 & -29\ 27\ 43.3 & -108.62 & -107.3 & -109.6 & -108.62 & 12.3\pm\ 0.8         & -0.45 & -0.49 & -0.48 \\
\rm NGC\ 6388 & 17\ 36\ 17.23 & -44\ 44\ 07.8 & 83.11 & 80.1   & --     & 83.11   & --                   & -0.55 & --    & -0.49 \\
\rm NGC\ 6522 & 18\ 03\ 34.02 & -30\ 02\ 02.3 & -15.23 & -21.1  & --     & -15.23  & 12.8\pm\ 1.0         & -1.34 & --    & -1.22 \\
\rm NGC\ 6528 & 18\ 04\ 49.64 & -30\ 03\ 22.6 & 211.86 & 206.6  & --     & 211.86  & 12.7\pm\ 1.0         & -0.11 & --    & -0.16 \\
\rm NGC\ 6553 & 18\ 09\ 17.60 & -25\ 54\ 31.3 & -0.27 & -3.2   & --     & -0.27   & 12.0\pm\ 2.0         & -0.18 & --    & -0.19 \\
\rm NGC\ 6558 & 18\ 10\ 17.60 & -31\ 45\ 50.0 & -195.12  & -197.2 & -192.6 & -195.12 & 12.3\pm\ 1.1         & -1.32 & -1.15 & -0.99 \\
\rm NGC\ 6569 & 18\ 13\ 38.80 & -31\ 49\ 36.8 & -49.83  & -28.1  & -49.7  & -49.83  & 12.8\pm\ 1.0^{\rm f} & -0.76 & -1.04 & -0.92 \\
\rm NGC\ 6626 & 18\ 24\ 32.81 & -24\ 52\ 11.2 & 11.11  & 17.0   & --     & --      & 12.8\pm\ 1.0         & -1.32 & --    & --    \\
\rm NGC\ 6656 & 18\ 36\ 23.94 & -23\ 54\ 17.1 & -148.72 & -146.3 & -147.4 & -148.72 & --                   & -1.70 & -1.75 & -1.70 \\
\rm NGC\ 6723 & 18\ 59\ 33.15 & -36\ 37\ 56.1 & -94.39 & -94.5  & --     & -94.39  & 12.6\pm\ 0.6         & -1.10 & --    & -1.02 \\
    \hline 
\end{array}
$$
\tablefoot{$^{\rm a}$ Center coordinates from the \cite{Harris1996} Catalog (2010 edition). $^{\rm b}$ \cite{BaumgardtRV}. $^{\rm c}$ \cite{capos_bulge}. $^{\rm d}$ \cite{Schiavon2024}. $^{\rm e}$ Ages from \cite{Bica24}, except marked differently. $^{\rm f}$ \cite{Saracino2019A6569}}
\end{table*}

\begin{table*}[t]
\caption{Summary of cluster properties as derived in this work.}
\label{tab:clusters}
$$
\begin{array}{lcrcccc}
    \hline \hline
Cluster & N_{members} & <RV_{\textrm{helio}}> & \sigma RV_{\textrm{helio}} & <[Fe/H]> & \sigma [Fe/H] & Classification^{\rm *} \\
    \hline 
\rm HP\ 1     & 3  & 37.91   & 3.90 & -1.20 & 0.11 & \rm B| B| B\\
\rm NGC\ 6093 & 7  & 10.14   & 4.29 & -1.59 & 0.08 & \rm LE| B| LE\\
\rm NGC\ 6171 & 19 & -32.31  & 3.49 & -0.93 & 0.07 & \rm B| B| D\\
\rm NGC\ 6266 & 6  & -76.25  & 11.25 & -1.03 & 0.05 & \rm B| B| B\\
\rm NGC\ 6304 & 11 & -108.61 & 4.41 & -0.51 & 0.08 & \rm B| B| D\\
\rm NGC\ 6388 & 12 & 79.99   & 4.71 & -0.51 & 0.10 & \rm B| B| LE\\
\rm NGC\ 6522 & 13 & -14.82  & 4.68 & -1.10 & 0.08 & \rm B| B| B\\
\rm NGC\ 6528 & 10 & 210.45  & 3.88 & -0.20 & 0.09 & \rm B| B| B\\
\rm NGC\ 6553 & 12 & -3.50   & 5.28 & -0.31 & 0.06 & \rm B| A| D\\
\rm NGC\ 6558 & 2  & -194.37 & 4.84 & -1.10 & 0.03 & \rm B| B| B\\
\rm NGC\ 6569 & 8  & -49.64  & 4.89 & -0.89 & 0.06 & \rm B| A| D\\
\rm NGC\ 6626 & 4  & 13.51   & 7.85 & -1.13 & 0.03 & \rm B| B| B\\
\rm NGC\ 6656 & 4  & -148.24 & 6.15 & -1.61 & 0.07 & \rm D| A| D\\
\rm NGC\ 6723 & 24 & -96.00  & 4.86 & -0.99 & 0.04 & \rm B| B| U\\
    \hline 
\end{array}
$$
\tablefoot{$^{\rm *}$ Classifications are compiled from the literature in the order of \cite{Massari2019}, \cite{Callingham2022}, and \cite{DeLeo2026}, denoted as B (bulge), D (disk), A (accreted), LE (low-energy), and U (unknown) according to their respective studies.}
\end{table*}

\subsection{Spectroscopic data}

We collected and analyzed the processed UVES slit spectra available from the European Southern Observatory (ESO) archive\footnote{\url{https://archive.eso.org/cms.html}}, along with  data obtained in multi-object spectroscopy (MOS) mode, where UVES is fed by the Fibre Large Array Multi Element Spectrograph \citep[FLAMES,][]{Pasquini2002}. The raw MOS data were reduced with the ESO Reflex pipeline \citep{ESOReflex}. To select the relevant objects, we used the 2010 version of the \cite{Harris1996} catalog to obtain a list of Galactic GCs and their center coordinates. We then downloaded archival spectra of stars located within a 20 arcmin radius around each GC. For more details on this procedure and the method used to identify the object associated with each spectrum, we refer to \cite{CHESSI}. 

We obtained and initially processed data for 395 stars in the field of 23 bulge GCs. The number of available archival spectra per cluster may be higher as, by default, we excluded those with a signal-to-noise ratio of S/N $<$ 10 prior to the analysis. After the membership selection and analysis, the sample was reduced to 135 member stars from 14 bulge GCs (see Table \ref{tab:clusters_summary}). We cross-matched all of them with external catalogs; however, erroneous identifications still remain possible in such crowded fields. For all the data we analyzed, we also list metadata (dates, program ID, and full file name)\footnote{To be published online on CDS: \url{https://cds.unistra.fr/}} that can be used to trace back the spectra in the ESO data archive. The S/N distribution of the data is shown in Fig.\ \ref{fig:SNR}.

To calculate the radial velocities (RVs) of the slit and MOS spectra, we used the method implemented in CHESS. It uses a cross-correlation of the target spectrum with different synthetic spectra templates, as summarized in \citet{CHESSI}. Heliocentric RVs (RV$_{\rm helio}$), used for membership analysis, were estimated separately for slit and MOS spectra. 

For the MOS data, we calculated the RVs for each individual exposure. The choice of co-adding multiple spectra was only considered for data with the same stellar ID in the header, obtained in the same observation program, and using the same UVES setup. This ensures that the co-added spectra belong to the same object and avoids possible mistakes made during our cross-matching with external catalogs for stellar identification. Furthermore, we first calculated the RV$_{\rm helio}$ for each spectrum of the same star using Astropy \citep{Astropy2022} functions and only co-added the ones that were within $\pm$0.5\ km~s$^{-1}$ of the mean RV$_{\rm helio}$.

For the slit spectra, when possible, the information available in the header was used to calculate RV$_{\textrm{helio}}$. We found three different cases. Some spectra contain the keyword ``specsys=HELIOCEN'' indicating that some pre-processing makes the measured RVs equal to RV$_{\textrm{helio}}$. For spectra in the format ``specsys=TOPOCENT'', there are two types of headers: those that include a keyword ``VRAD\_HELIOCOR'' and those that do not. This keyword contains the value of the calculated heliocentric corrections. When available, we added these corrections to our RV values. In the cases where no correction was provided, we calculated them for the mid-observation time in a manner similar to that done for the MOS data, using Astropy functions. The mean values of RV$_{\textrm{helio}}$ for the clusters are given in Table \ref{tab:clusters}.

\subsection{Membership determination} \label{sec:members}

To evaluate whether the stars belong to a cluster, we used their RV$_{\textrm{helio}}$ and their metallicities, which were derived according to the process discussed in Section \ref{sec:method}. In the first step, we considered the object to be a field star if its RV$_{\textrm{helio}}$ differed by more than $\pm 15$ km s$^{-1}$ (typical internal dispersion in GCs is $\sim 5\ \rm{km~s}^{-1}$) from the value reported in the literature. Our collection of mean RV$_{\textrm{helio}}$ values includes results from \cite{BaumgardtRV}, \cite{capos_bulge}, \cite{Schiavon2024}, and \cite{Harris1996}, in this order. 

However, because the bulge line of sight is complex, some non-member stars may end up sharing similar RV$_{\textrm{helio}}$ with the cluster. Therefore, we also used a comparison of the metallicities to exclude such stars. As the source of values, we prioritized the following catalogs in this order: the bulge Cluster APOgee Survey \citep[CAPOS,][]{capos_bulge} and APOGEE Value Added Catalogue of \citet[APOGEE VAC,][]{Schiavon2024}, \citet{Bica24}, and \citet{Harris1996}. Both CAPOS and APOGEE VAC are recent works that share a similar metallicity scale, whereas values from the Harris catalog may be out-dated. For example, for NGC 6316, the Harris catalog lists a value that is about 0.4 dex higher than those two catalogs. This is mainly because the Harris catalog is not homogeneous in terms of its methodology and data. 

We first evaluated the difference between the maximum and minimum metallicities among the literature values. If the difference exceeded 0.2 dex, we accepted as members the stars with [Fe/H] within $\pm 0.35$ dex of the chosen literature value. When the literature values agreed to within 0.2 dex, we accepted as member stars those within $\pm 0.20$ dex of the chosen literature metallicity. Applying such a selection filter is justified only because our sample does not include stars belonging to clusters that are known to exhibit a large intrinsic metallicity spread, the so-called iron complex clusters, such as Liller 1 \citep{Fanelli2024, Chiappino2026} and Terzan 5 \citep{Origlia2013, Massari2014}. In our sample, only NGC 6656 (M22) is known to have complex chemistry with a spread of metallicity values \citep{Marino2009, Marino2011, Alves-Brito2012}. This cluster will be studied in more detail in a forthcoming paper. Here, we kept it in the sample as a tracer of the very metal-poor population in the direction of the bulge.

As a consistency check, we compared our sample with the membership probabilities of \cite{Vasiliev2021GaiaClusters}. Only ten of the 135 selected stars have membership probabilities lower than 70\%\ in their work. We did not exclude those ten stars since they have the same RV and metallicity values as their parent clusters. The number of member stars per cluster is given in Table \ref{tab:clusters}.

\section{Methodology}\label{sec:method}

With the CHESS pipeline, our aim has been to obtain results of high precision and accuracy. When possible, we perform a line-by-line differential analysis (for precision) of sample stars against similar reference stars that have accurate atmospheric parameters. We identified similar stars directly from their spectra, before the radiative transfer analysis, using the dimensionality reduction algorithm called t-SNE. In this ``similarity analysis,'' the pixels in a spectrum are treated as dimensions. Each stellar spectrum becomes a point in a multidimensional space where similar stars cluster together. Using t-SNE, we obtained a 2D projection map of this space that can be used to identify stars that are similar to each reference object within $\pm$ 200 K, $\pm$ 0.3 dex, and $\pm$ 0.2 dex in effective temperatures, surface gravities, and metallicities \citep[see][for details on this method]{CHESSI}. 

However, in the case of the current sample, we found two challenges in fully implementing this approach. First, only a small number of our targets (11 stars) matched reference objects because of the small number of metal-poor low surface gravity reference stars. Second, our tests indicated that the low S/N values of our spectra restricted the benefit in precision coming from a differential analysis. We therefore modified the methodology, as detailed below, to exploit the reference stars to quantify and correct potential systematic biases inherent to the implemented spectroscopic analysis. This approach ensures accuracy, putting the results on the same scale of the reference stars.

\subsection{Atmospheric parameter determinations}

In our adapted method, we first performed a traditional spectroscopic analysis based on the excitation and ionization equilibrium of Fe lines. We used q2 \citep{Ramirez14}, a wrapper that automates the use of the  MOOG\footnote{\url{https://www.as.utexas.edu/~chris/moog.html}} \citep{moog} line analysis code. The one-dimensional MARCS \citep{Marcsmodels} model atmospheres were adopted. Our line list is a compilation of selected \ion{Fe}{i} and \ion{Fe}{ii} lines from different sources \citep{Melendez2009, Melendez2014, Ramirez14, Barbuy2021, Heiter21, Shi2025}. The code REvIEW \citep{McKenzie22}, with some customization of the output formats and plotting functions,  was used to measure the equivalent widths (EWs) of the normalized spectra. The normalization is carried out using SUPPNet \citep{SUPPNet}, as described in \cite{CHESSI}. In the second step, we implemented empirical corrections to remove biases and bring the atmospheric parameters to an accurate scale.  

To find the biases that are intrinsic to our method of parameter estimation, we also performed the traditional spectroscopic analysis for a series of reference stars with well determined, accurate parameters spanning a larger parameter space than the one relevant for the bulge GC stars. We used the following as reference sample: the Sun, TITANS I and TITANS II metal-poor stars \citep{Giribaldi2021, Giribaldi2023}, \textit{Gaia}-ESO Survey K2 sample of giants \citep{Worley2020}, and \textit{Gaia} benchmark stars \citep{Soubiran2024, Casamiquela2026}. The effective temperatures (\teff) of the TITANS were derived via 3D NLTE modeling of the H$\alpha$ line and show an accuracy of 1\% compared to \teff~values based on interferometry. The surface gravities (\logg) were derived with evolutionary tracks and parallaxes for the TITANS I dwarf stars \citep{Giribaldi2021} and with the fitting of \ion{Mg}{i} b triplet lines for the TITANS II giants \citep{Giribaldi2023}, giving comparable results to those obtained using asteroseismology. The parameters for the \textit{Gaia}-ESO Survey K2 stars were derived by an iterative spectroscopic-asteroseismic analysis \citep{Worley2020}. The analysis of the \textit{Gaia} benchmark stars was done using angular diameters, bolometric fluxes, parallaxes, and evolutionary models \citep{Soubiran2024, Casamiquela2026}. 

\begin{figure}
    \centering
    \includegraphics[width=\linewidth]{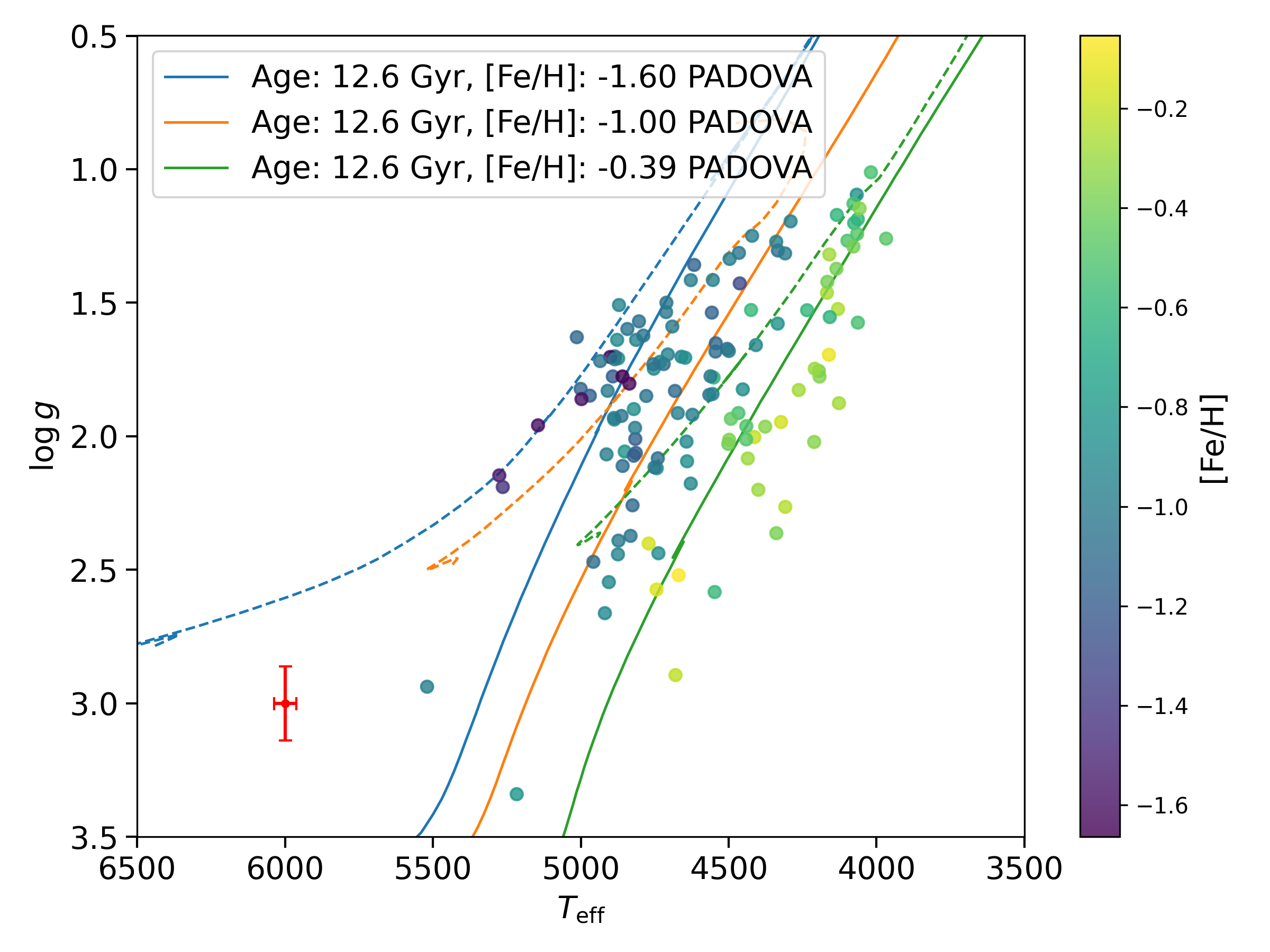}
    \caption{Kiel diagram for the final sample with PADOVA isochrones \citep{Parsec, Parsec2} for three different metallicities (see inset) all with an age of 12.6 Gyr. Solid and dashed lines represent isochrones before and after the tip of the RGB, respectively. The stars are color-coded by metallicity, as shown in the side bar. A mean error bar is shown in the bottom left ($\pm$38 K and $\pm$0.14 dex for \teff\ and \logg, respectively).}
    \label{fig:kiel_sample}
\end{figure}

To determine the bias correction that needs to be applied to our sample, we computed the difference (i.e., bias) between the derived and reference parameters. We then proceeded to fit the differences in each parameter individually as a quadratic function of the other three parameters, \teff, \logg, and [Fe/H] (including the cross terms) and used these functions to estimate and correct the biases that affect our results for the rest of the sample. More details on this calculation will be published in a forthcoming paper (Smiljanic et al., in preparation). The functions used here are given in Appendix \ref{ap:bias}. A Kiel diagram with the final (bias corrected) atmospheric parameters is shown in Fig. \ref{fig:kiel_sample}. The figure shows that our corrected parameters agree well with the expected position of red giants in the metallicity range of our GCs, validating our approach.

\subsection{Elemental abundances}

We derived the elemental abundances based on a spectral synthesis using the TSFitPy \citep{Storm2023} wrapper of the Turbospectrum \citep{Alvarez1998, Plez2012} code. We mainly used the atomic and molecular data from the \textit{Gaia}-ESO survey \citep{Heiter21} to synthesize the spectra; however, for the Ba and Eu lines ,we adopted the hyperfine structure from \cite{daSilva_Smiljanic_2025}. The lines used in this analysis are summarized in Appendix \ref{apx:elemental_lines}. For the solar scale, we used \cite{Grevesse2007}. Abundances were first determined in LTE for Li, O, Na, Mg, Al, Si, Ca, Sc, Y, Zr, Ba, Ce, Nd, and Eu. The functionality of taking into account deviations from the local thermodynamic equilibrium (i.e., so-called non-LTE or NLTE effects) was additionally employed for Na \citep{Ezzeddine2018}, Mg \citep{Bergemann2017}, Al \citep{Ezzeddine2018}, Si \citep{Magg2022}, Ca \citep{Mashonkina2017, Semenova2020}, Y \citep{Storm2023, Storm2024}, Ba \citep{Gallagher2020}, and Eu \citep{Storm2024}; see \citet{Gerber2023}. The LTE and NLTE abundances have been published online on CDS \footnote{https://cds.unistra.fr/}.

For individual line abundances, we excluded TSFitPy solutions that had error flags, chi-square values of $>$ 4, and EW $<$ 10 m\AA. Moreover, we removed solutions that converged but needed the macroturbulence velocity ($V_{\rm macro}$) with values $> 10$\ km s$^{-1}$. To fit the broadening of each line separately, we opted to leave $V_{\rm macro}$ as the only free parameter and input fixed values of spectral resolution (with appropriate values for each spectrum) and rotational velocity (at 1.2 km s$^{-1}$). This is justified as at the typical UVES resolution and if the stars are not fast rotators, all three broadening effects are degenerate. A large $V_{\rm macro}$, which in this case is a combined broadening and not only the macroturbulence, was taken as an indication that a particular line had some peculiarity preventing good quality abundances from being determined. After cleaning the sample with these filters, we already reached a line-to-line scatter of less than 0.2 dex for elemental abundances in most of the sample. We further investigated line fits visually in the following cases: if only one line of the element was measured for a given star, if the standard deviation of the abundances was greater than 0.2 dex, or if the measured abundance for a given element was much higher than the sample average (e.g., Al, where it can reach $>$0.7 dex). We derived parameters for 135 stars from 14 bulge GCs and abundances for 114 stars from 13 GCs (from data with S/N $>20$, see Appendix \ref{apx:elements} for the mean cluster abundances).

\begin{figure}
    \centering
    \includegraphics[width=\linewidth]{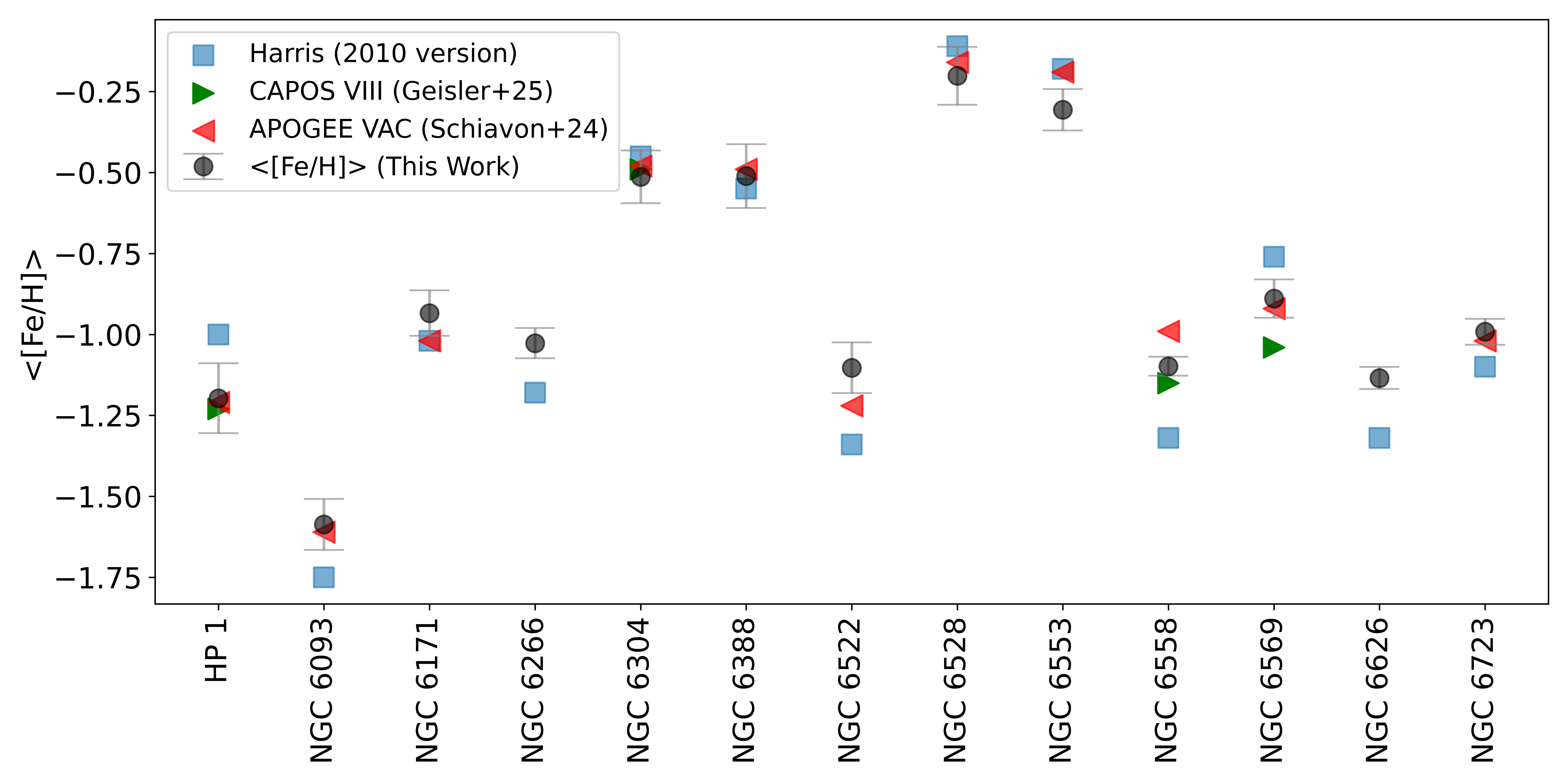}
    \caption{Comparison of metallicities for the analyzed clusters.}
    \label{fig:cluster_metal_comp}
\end{figure}

\subsection{Uncertainties in the parameters and abundances}\label{sec:errors}

The errors in the parameters were calculated with the q2 code and were based on the trends needed to achieve excitation and ionization equilibrium of the Fe lines. Individual errors of derived parameters are part of the online tables. The 90th percentile of the error distribution is 66 K, 0.22 dex, 0.08 dex, and 0.14 km s$^{-1}$ for \teff, \logg, [Fe/H], and microturbulence (\vmicro), respectively. 

We used these values to find the uncertainties of the abundances. For this purpose, we chose three representative stars distributed along the metallicity range of the sample. The mean changes in the derived abundances typically range between 0.10 and 0.16 dex, as summarized in Appendix ~\ref{apx:errors}.

\section{Discussion}\label{sec:discussion}

\subsection{Metallicity values}

We compared the metallicities obtained in this work with values from APOGEE VAC \citep{Schiavon2024}, CAPOS VIII \citep{capos_bulge}, and \cite{Harris1996} (see Fig. ~\ref{fig:cluster_metal_comp}). While the Harris catalog reports metallicity values for each GC, we note that APOGEE VAC and CAPOS VIII reported metallicities for twelve and five GCs in common with our work, respectively. NGC 6266 and NGC 6626 have metallicity values reported only in this work and in the Harris catalog. As a consequence of our membership selection, which rules out stars with metallicities different by $\pm$0.35 dex compared to the literature, the general agreement in the metallicity among stars in a cluster is very good and with small standard deviation. Here, HP 1 is the case with the largest spread in the metallicity, but with a small number of members (only three). NGC 6723, on the other hand, has the highest number of member stars and displays a very small standard deviation ([Fe/H] = $-$0.99 $\pm$ 0.04 dex from 24 stars). The very good agreement between our results and those from homogeneous spectroscopic surveys (APOGEE and CAPOS) demonstrates the robustness of our analysis, even for data with relatively low S/N values.

The bulge GCs are known to show a bimodal metallicity distribution \citep{Bica16, Bica24}, with peaks at [Fe/H] $\sim$ $-$1.10 and $-$0.50. This seems to reflect on our results, although our sample size is relatively small (N$_{cluster}$=14) to draw any firm conclusion. Eight of the GCs we analyzed have [Fe/H] values between $-$1.20 and $-$0.80 while four have [Fe/H] between $-$0.6 and $-$0.20. A similar multi-modal metallicity distribution has also been reported for bulge field stars \citep{GIBS2017, Rojas-Arriagada2017, Rojas-Arriagada2020}. There is broad agreement that the metal-rich stars are linked to the secular evolution of the disk into an X-shaped bar. The metal-intermediate and metal-poor stars may have formed in gas-rich environments that underwent rapid star formation. The metal-poor tail of the bulge, around [Fe/H] = $-$1.3, has been identified in several studies \citep{Rojas-Arriagada2014, Ness2013} and is associated with an old bulge population \citep{Bica16}.

The origins of bulge GCs remain a topic of debate, with recent improvements in dynamic studies driven by \textit{Gaia} data. The majority of our clusters are identified as bulge GCs in three different studies \citep[namely HP 1, NGC 6266, NGC 6522, NGC 6528, NGC 6558, NGC 6626][]{Massari2019, Callingham2022, DeLeo2026}. NGC 6723 is identified as a bulge cluster in \cite{Massari2019} and \cite{Callingham2022}, but of unknown classification in \cite{DeLeo2026}. NGC 6656 is an ex situ cluster \citep{PerezVillegas2020, capos_bulge} with complex chemistry \citep[see][]{McKenzie22, capos_bulge}. Whether NGC 6093 is a true bulge GC is debated \citep{capos_bulge} because its [Fe/H] is below the $-$1.5 dex threshold used by \cite{Bica16} to identify halo intruders. Apart from NGC 6656 and NGC 6093, the other clusters in the sample are classified as bulge or disk clusters in these studies.

\subsection{Serendipitous discovery of a new Li-rich giant}

We were able to derive the Li abundance for only two stars. This is caused by a combination of the low S/N and the fact that Li is usually depleted in giants, which makes its line weak. We found A(Li) = 1.22 for the star \textit{Gaia} DR3 4330462795522798592 in NGC 6171. We could detect the line even though the spectrum has a low S/N ($\sim$ 24). Comparing this star with the sample of Li-rich giants assembled in \cite{Sanna2020}, we conclude that it is a typical giant with normal Li abundance. 

\begin{figure}
    \centering
    \includegraphics[width=0.9\linewidth]{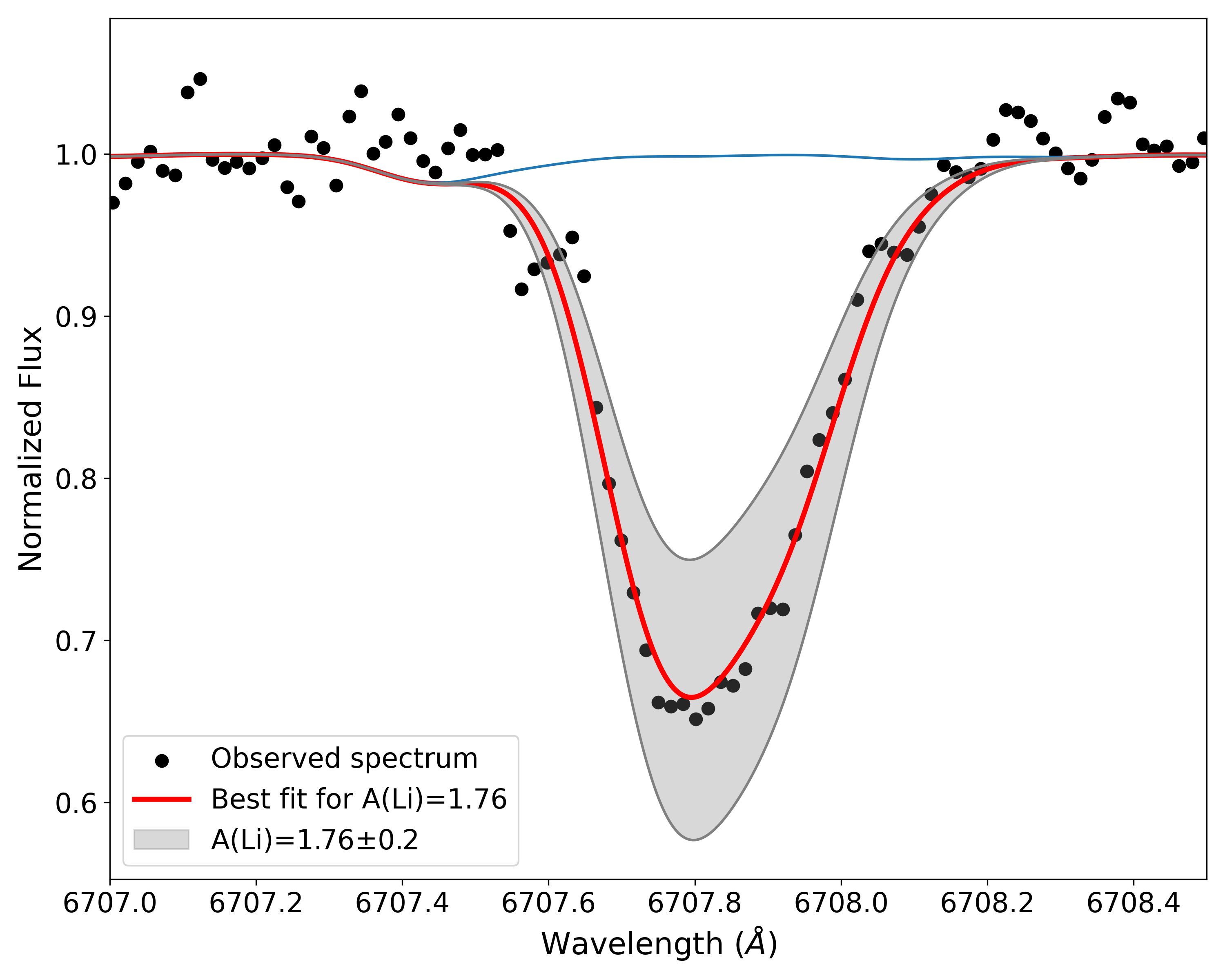}
    \caption{Li line for the Li-rich giant star in NGC 6723.}
    \label{fig:li_fit}
\end{figure}

The second giant with detected Li turned out to be the first Li-rich giant star discovered in NGC 6723. Lithium-rich giants are very rare in GCs, only 16 of them have been reported in total thus far \citep{Sanna2020, Aguilera2022}. In its observation block, the star is named 1018221 and was observed as part of the ESO Program 193.D-0232. According to our target identification, this star is \textit{Gaia} DR3 6730905075064875648. We derived for this giant (\teff, \logg, [Fe/H], \vmicro) =  (4913 K, 2.07 dex, $-$0.98 dex, 1.62 km s$^{-1}$). Its Li line is shown in Fig.\ \ref{fig:li_fit}. We find A(Li) = 1.76 in LTE, which puts it in the main group of Li-rich stars in GCs \citep{Sanna2020}. The 3D NLTE correction calculated using the Breidablik \citep{Wang2021, Wang2024} code is $-$0.08 dex. According to \cite{Vasiliev2021GaiaClusters}, it has 100\% cluster membership probability.

The abundance of Na in the Li-rich giant does not show significant enrichment, suggesting that this star belongs to the first generation of the cluster ([Na/Fe]$_{LTE}$ = 0.07, placing it on the limit between first- and second-generation stars for this cluster if we adopt the [Na/Fe]$_{\rm min}$ + 0.3 criteria from \citealt{Carretta2009a}). The other abundances agree with the cluster average except for Ba. It has [Ba/Fe]$_{\rm NLTE}$ = $-$0.28 while the cluster average is [Ba/Fe]$_{NLTE}$ = +0.26 (excluding the Li-rich star). 

\begin{figure}
    \centering
    \includegraphics[width=\linewidth]{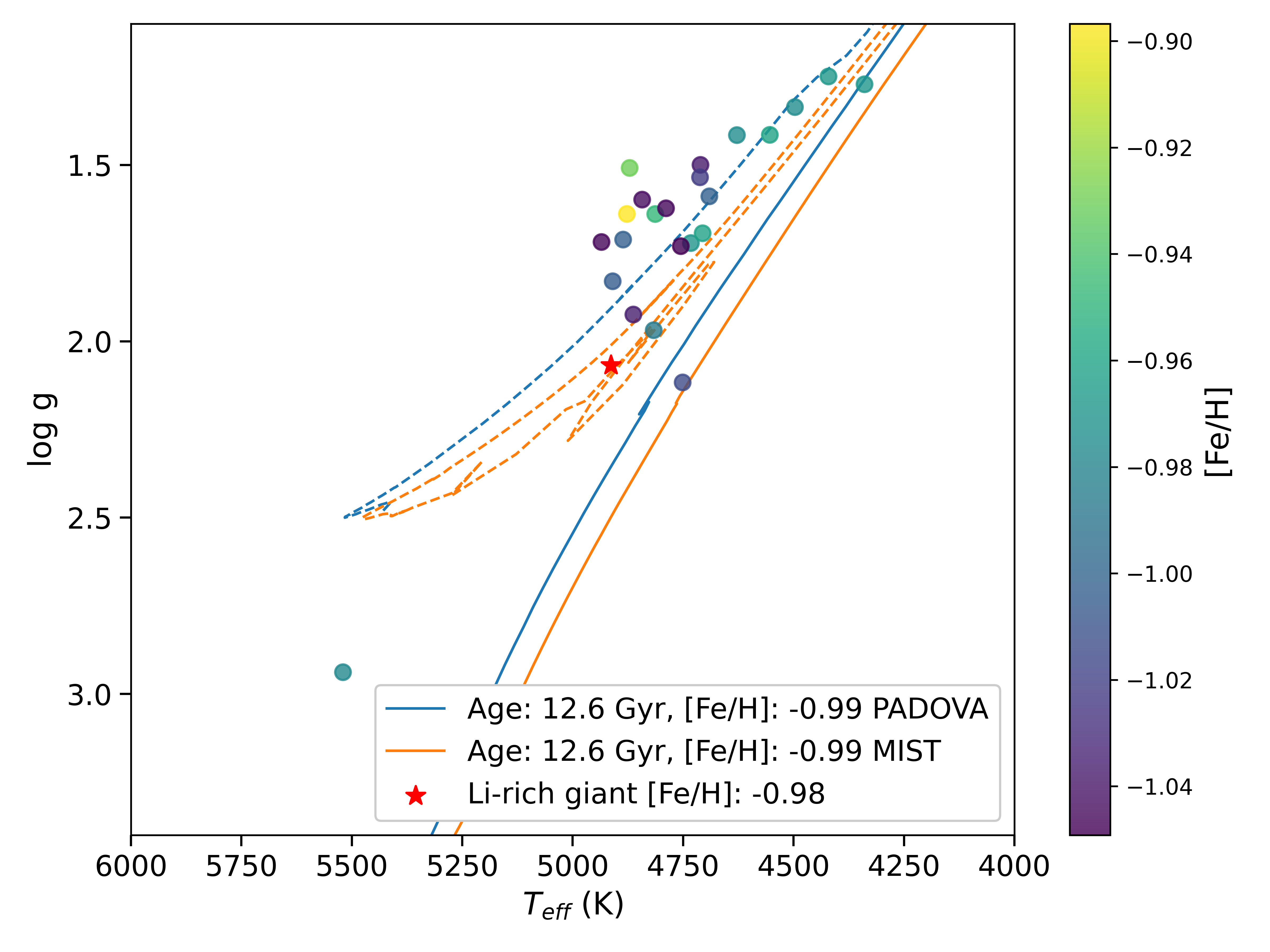}
    \caption{Kiel diagram of NGC 6723 with MIST and PADOVA isochrones. The Li-rich giant is shown as a red star. The remaining stars are color coded by metallicity. Lines are as in Fig ~\ref{fig:kiel_sample}. The isochrones have the mean cluster metallicity derived here and age from \cite{Bica24}.}
    \label{fig:kiel_li_rich}
\end{figure}

The location of the star on the Kiel diagram (Fig.~\ref{fig:kiel_li_rich}) is consistent with the core-He burning phase. This agrees with the hypothesis that most Li-rich giant stars are red clump stars \citep{Casey2019, DeepakLambert2021, Yan2021}. Possible explanations invoked to explain Li-rich stars include fresh internal Li production, mass transfer from a binary companion that went through the asymptotic giant branch (AGB) phase, and planet engulfment \citep[see][and references therein]{Pasquini2014, Sanna2020}. We only have spectra obtained in one epoch and therefore cannot investigate RV variation. The low abundance of Ba seems to argue against mass transfer from an AGB companion \citep[as in the star discussed in][]{Sanna2020}. For now, intrinsic Li enrichment seems to be the most likely explanation. To further study the issue, higher quality data are needed to allow the investigation of elemental abundances that can help to better understand the star in the context of stellar evolution and multiple populations (e.g. abundances of C, N, O, and Al) including also the $^{12}$C/$^{13}$C ratio as an evolutionary stage diagnostics.

\subsection{Light element abundances and (anti-)correlations}

\begin{figure}
    \centering
    \includegraphics[width=\linewidth]{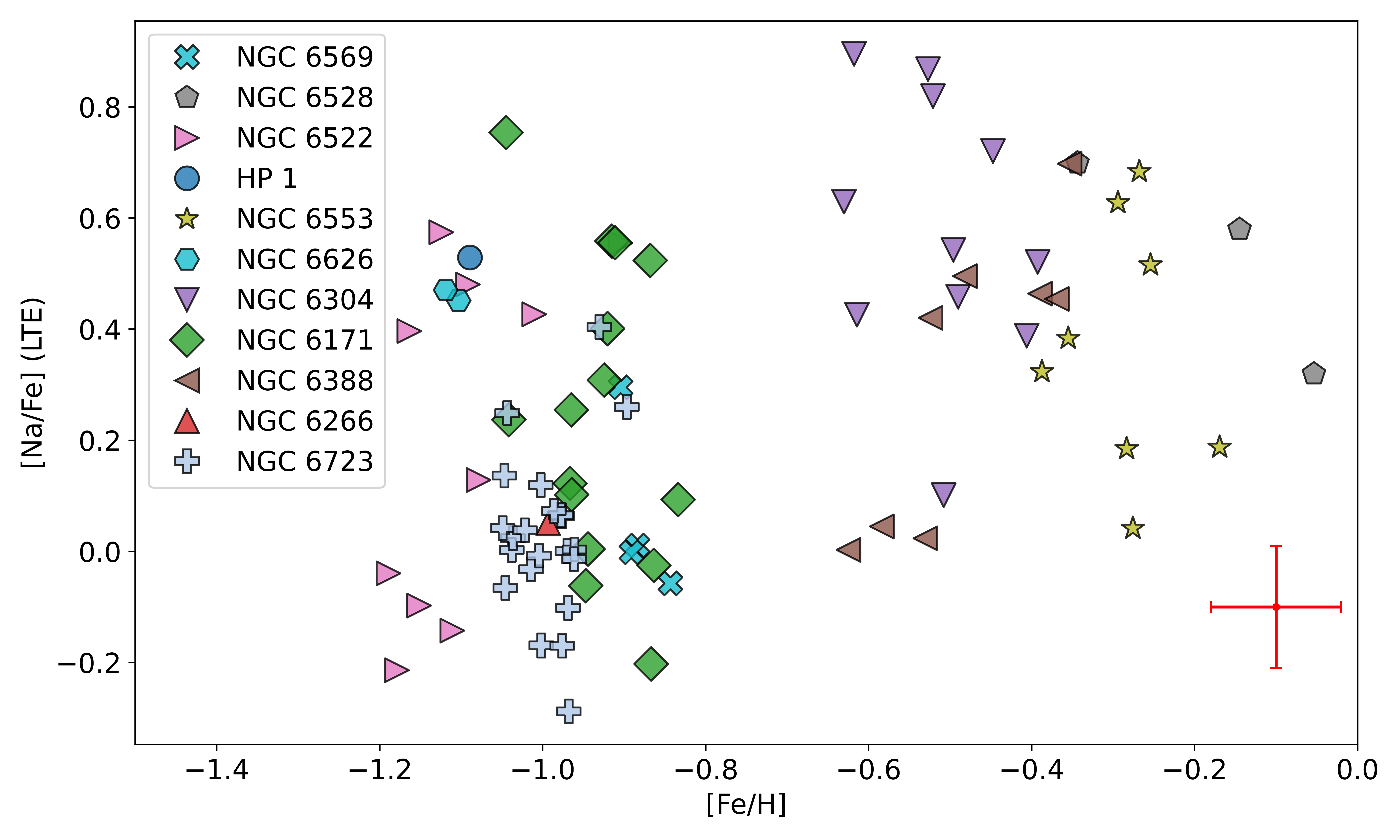}
    \caption{Na abundances vs. metallicity of individual stars. It is visible that our sample consists of both first-generation (Na-poor) and second-generation (Na-rich) stars if we use [Na/Fe]$_{min}$ and [Na/Fe]$_{min}$ + 0.3 criteria \citep{Carretta2009a} for a given cluster.}
    \label{fig:Na_abund}
\end{figure}

\begin{figure}
    \centering
    \includegraphics[width=\linewidth]{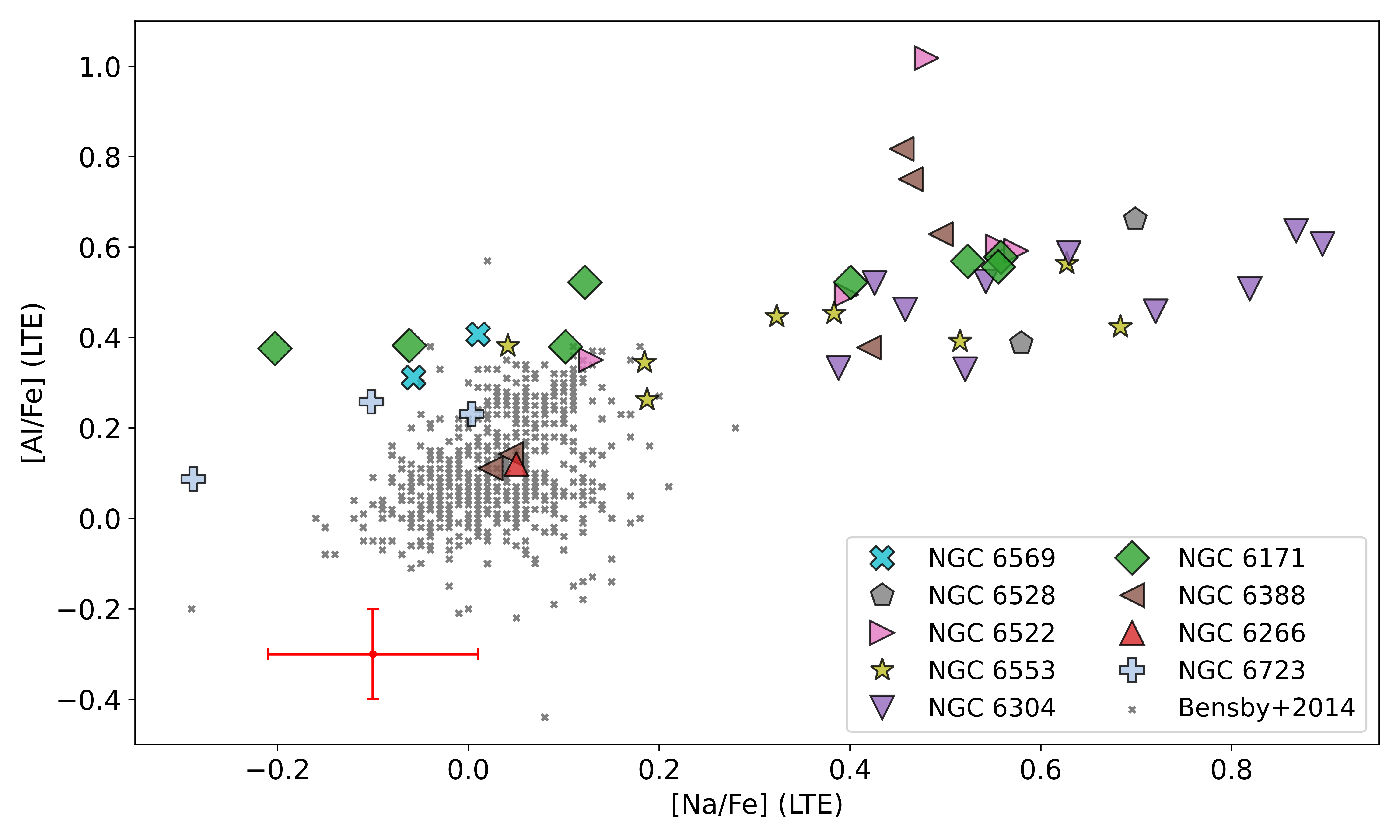}
    \caption{Correlation between Na and Al abundances. Clear separation is visible for different generations of stars. Gray points show stars from solar neighborhood \citep{Bensby2014}.}
    \label{fig:Na_Al}
\end{figure}

\begin{figure*}
    \sidecaption
    \includegraphics[width=0.7\linewidth]{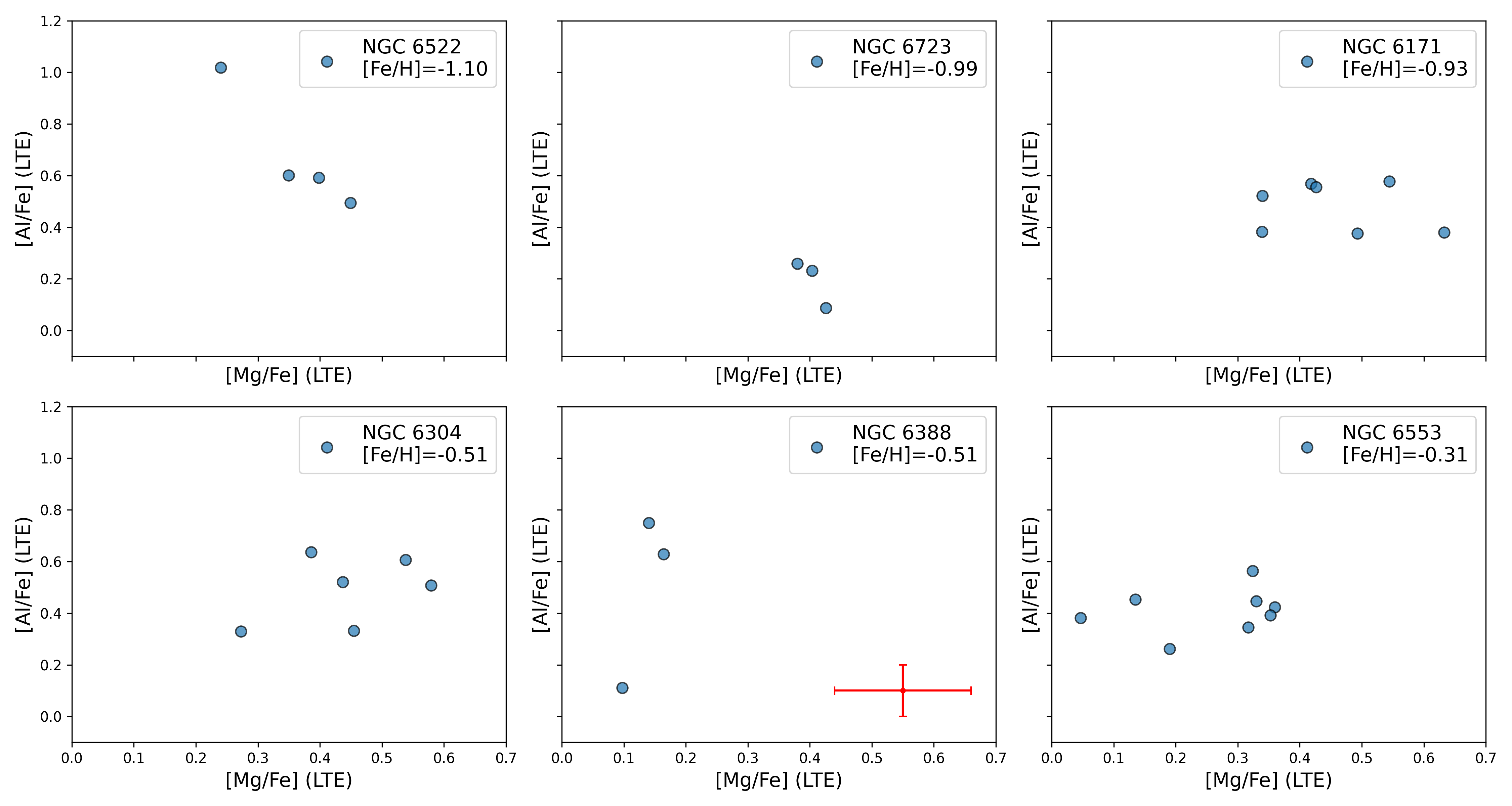}
    \caption{Trends between [Mg/Fe] vs. [Al/Fe] for the six GCs where these elements were derived for at least three stars. The GCs are ordered by increasing metallicity from the top-left to the bottom-right.}
    \label{fig:MgAl}
\end{figure*}

We could only report oxygen abundances derived from the 6300 \AA\ forbidden line for two stars. This was caused by a combination of low S/N and telluric lines that affected the line in many spectra. The features used to derive the C and N abundances were too weak and could not be used. We derived Na and Al abundances in both LTE and NLTE. We observe abundance spreads from the first- and second-generation stars  in the clusters NGC 6171, NGC 6304, NGC 6388, NGC 6522, NGC 6553, and NGC 6723 (Fig.~\ref{fig:Na_abund}\footnote{We plotted our LTE abundances in Figures from \ref{fig:Na_abund} to \ref{fig:Zr} for a more accurate comparison with the literature LTE abundances that are shown.}). All these clusters have stars with Na abundances outside the range between [Na/Fe]$_{\rm min}$ and [Na/Fe]$_{\rm min}$ + 0.3, following \citet{Carretta2009a}. Our remaining clusters do not have samples large enough to investigate this issue, especially considering the lack of O abundances. Figure \ref{fig:Na_Al} shows how the Na and Al abundances correlate and display a division caused by the different stellar generations. The correlation remains even if we select only the results for data with S/N $>$ 30 or S/N $>$ 40, suggesting that we are not seeing a scatter simply related to S/N. However, in the S/N $>$ 40 case, the sample is strongly reduced to eight stars with Na and Al, but still includes three stars from NGC 6171 and two stars from NGC 6388 showing abundance spreads. Spreads in both Na and Al need first-generation polluters that can achieve high temperatures to activate the NeNa and MgAl cycles \citep[see e.g.][]{Prantzos2007}.

The anti-correlation between Mg and Al is another sign of such high temperature cycles sometimes observed in GCs. The magnitude of this anti-correlation depends on the mass and metallicity of the cluster \citep{Pancino2017, Meszaros2020}. It was challenging to derive Al abundances for most of our sample, resulting in a limited number of stars having reported values (47 stars, with 35 of them having Mg abundances). Only six clusters have more than three stars with these abundances, enabling an investigation of the Mg-Al anti-correlation (Fig.\ \ref{fig:MgAl}). Among these clusters, the [Mg/Fe] versus [Al/Fe] plot shows a spread in Al only for NGC 6522 and NGC 6388. In NGC 6522, we can clearly see the spread in Al with a possible anti-correlation pattern with Mg. We note that \cite{Barbuy2021} instead found an Mg-Al correlation in their analysis of this cluster. For NGC 6388, \cite{Meszaros2020} did not detect any Al spread in their analysis, although they emphasized the need for a larger sample size. On the other hand, \cite{Carretta2007} and \cite{Carretta2023} concluded that Mg and Al are anti-correlated in NGC 6388. With only three stars having both Mg and Al abundances at our disposal, we were not able to confirm or reject the existence of an anti-correlation. Interestingly, NGC 6522 and NGC 6388 are in different metallicity peaks of the bulge GC distribution with [Fe/H] = $-$1.10 and $-$0.51, respectively.

\subsection{Abundances of alpha elements and scandium}

Proton-capture reactions at very high temperatures can, in principle, affect the abundances of elements heavier than Al and Mg, such as Si \citep{Ventura2012}. Some GCs with possible spreads in Ca and Sc, likely showing signs of such high-temperature reactions, have been reported \citep[see][and references therein]{CarrettaBragaglia2021}. For these reasons, we decided to include Ca, Si, and Sc in our study. We did not detect any significant spreads within GCs for these elements and, therefore, discuss them here only in the general context of the chemical evolution of the bulge.

\begin{figure*}
    \centering
    \includegraphics[width=\linewidth]{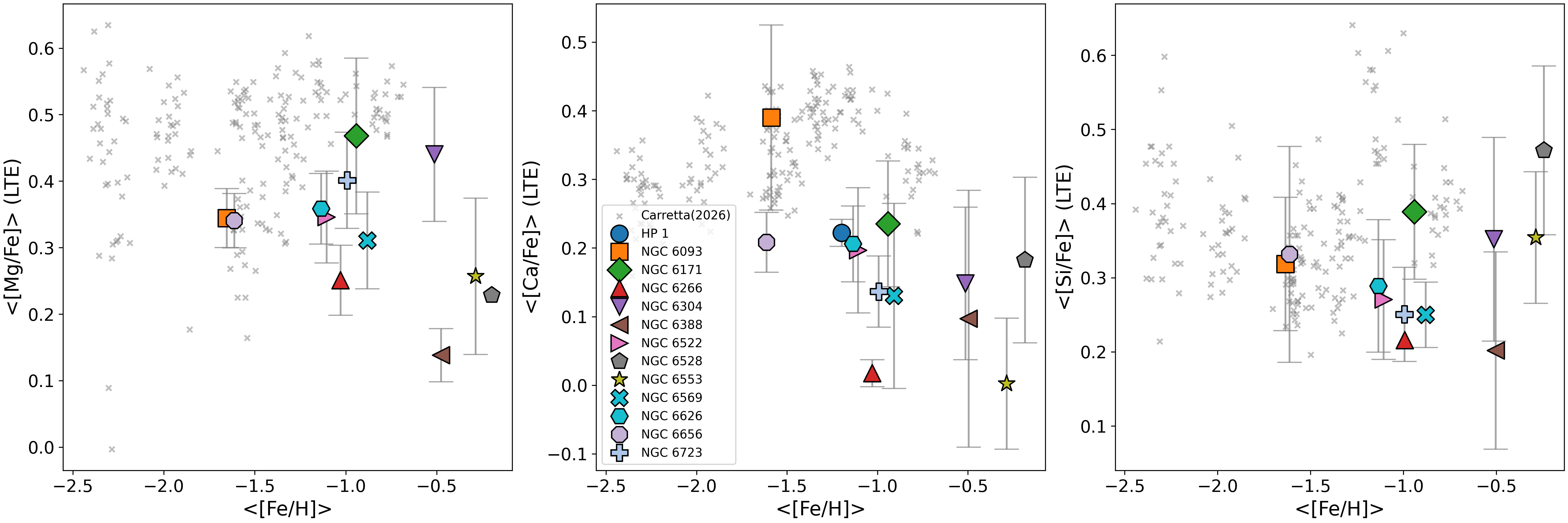}
    \caption{Element over iron ratios as a function of metallicity for the $\alpha$-elements derived in this work. Mean cluster abundances and their standard deviation are plotted. Symbols described in the middle panel are in common with other plots in the paper. Gray points shows abundances from \cite{Carretta2026b} and metallicities from \cite{Carretta2009a}.}
    \label{fig:alpha_indv}
\end{figure*}

\begin{figure}
    \centering
    \includegraphics[width=\linewidth]{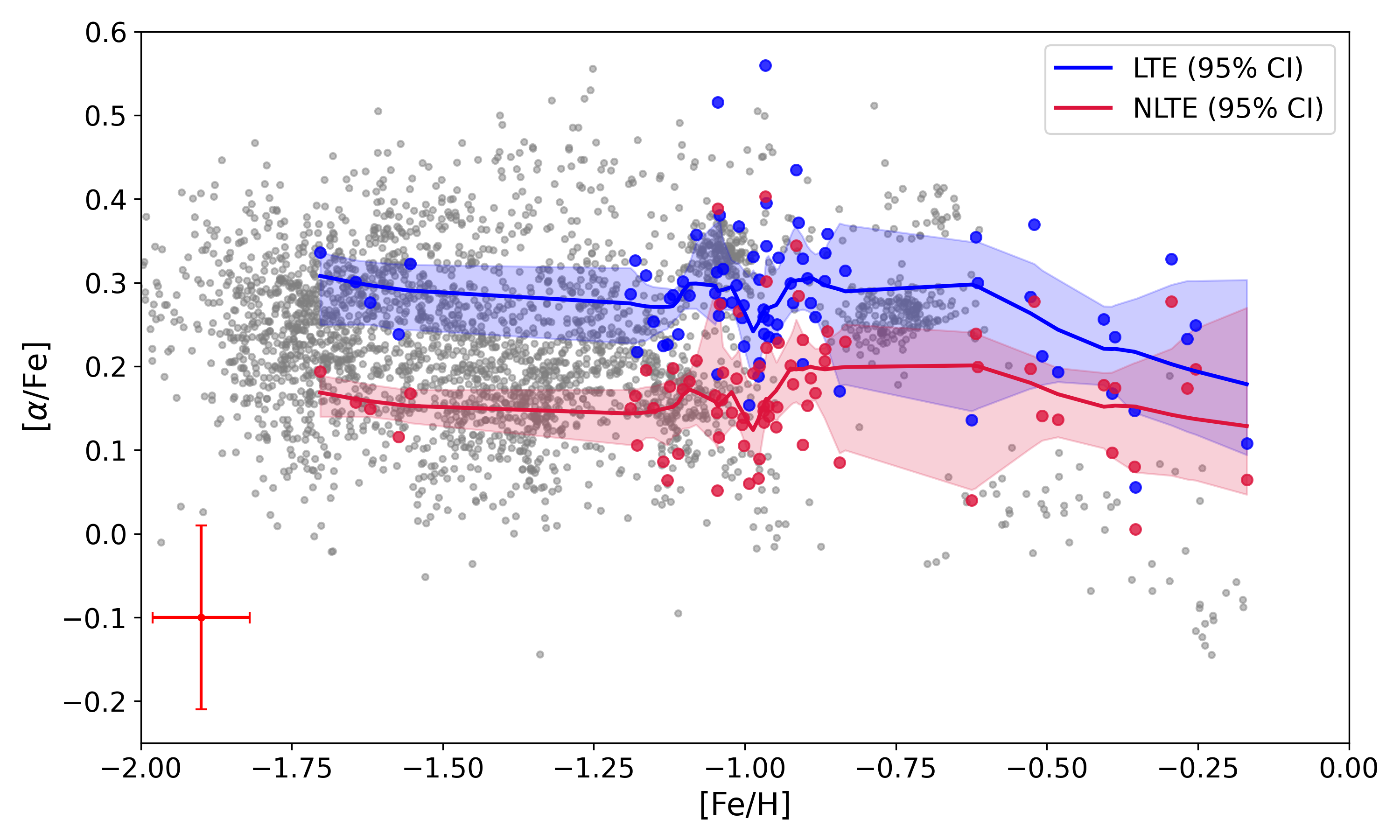}
    \caption{Evolution of a mean $\alpha$-element abundance with metallicity. The lines are locally weighted scatter plot smoothing (LOWESS) functions used to represent how the abundances change with metallicity. LOWESS is a non-parametric regression method to fit a smooth curve over the plotted data. The gray points are globular cluster stars from APOGEE VAC \citep{Schiavon2024}.}
    \label{fig:alpha}
\end{figure}

In Fig. ~\ref{fig:alpha_indv}, we show the mean cluster abundances of three $\alpha$-elements (Mg, Ca, and Si) versus metallicities, compared to the abundances of the comprehensive spectroscopic study of stars in GCs by \citet{Carretta2009a, Carretta2009b, Carretta2026a, Carretta2026b}. Their sample mostly consists of stars in halo globular clusters observed with GIRAFFE but with UVES data for some stars. As can be seen, our two very metal-poor clusters (most likely halo intruders) are shown to follow the literature trend for all three $\alpha$ elements. However, for the bulge GC sample, the Mg and Ca abundances are lower than the halo clusters in the same metallicity range ([Fe/H] $\gtrsim$ $-$1.0). Some difference in Si was also observed, although it is less pronounced than for Mg and Ca. In Figure ~\ref{fig:alpha}, we compare the general evolution of [$\alpha$/Fe]\footnote{Where $\alpha$ is an average of Mg, Ca, and Si abundances.} against [Fe/H] in our clusters with the sample of APOGEE VAC \citep{Schiavon2024}. For [Fe/H] $\lesssim$ $-$0.7 dex, our LTE and NLTE [$\alpha$/Fe] ratios have values of $>$ 0.0, but with a plateau that is lower than that seen in APOGEE VAC (where Mg and Ca are in NLTE, and Si in LTE). The mean [$\alpha$/Fe] decreases towards higher metallicities starting somewhere between [Fe/H] = $-$0.75 and $-$0.5. The decrease is stronger in the APOGEE VAC clusters, so our bulge GCs in this range have higher [$\alpha$/Fe] than seen in APOGEE VAC. This probably reflects the more intense star formation in the bulge regions. Interestingly, this metallicity range is the least broadly sampled in both studies. 

An interesting result is a difference in Mg between NGC 6388 and NGC 6304, which are characterized by just about the same metallicity ([Fe/H] $\sim$ $-$0.5). We found for NGC 6388 and NGC 6304, [Mg/Fe] = 0.14 $\pm$ 0.04 (mean and standard deviation based on five stars) and 0.44 $\pm$ 0.1 (mean and standard deviation based on seven stars), respectively. \cite{Carretta2023} reported <[Mg/Fe]> = 0.21 $\pm$ 0.05 for NGC 6388, which is slightly different, but still in agreement with our results within the uncertainties. For NGC 6304, \cite{capos_ngc6304} derived a lower abundance than ours (<[Mg/Fe]> = 0.29 $\pm$ 0.02). If confirmed, the difference that we detected could indicate distinct origins for the two clusters. NGC 6388 shows similar $\alpha$ element abundances to field stars from the disk and bulge, as discussed in detail in \cite{Carretta2023}. The Mg value of NGC 6304 would be closer to thick-disk or halo stars. However, since we made use of archival data, we cannot exclude that for NGC 6388, our sample preferentially includes second population stars with low Mg. A detailed comparison of this and other abundances from better data would be important to confirm and expand this finding. 

\begin{figure}
    \centering
    \includegraphics[width=\linewidth]{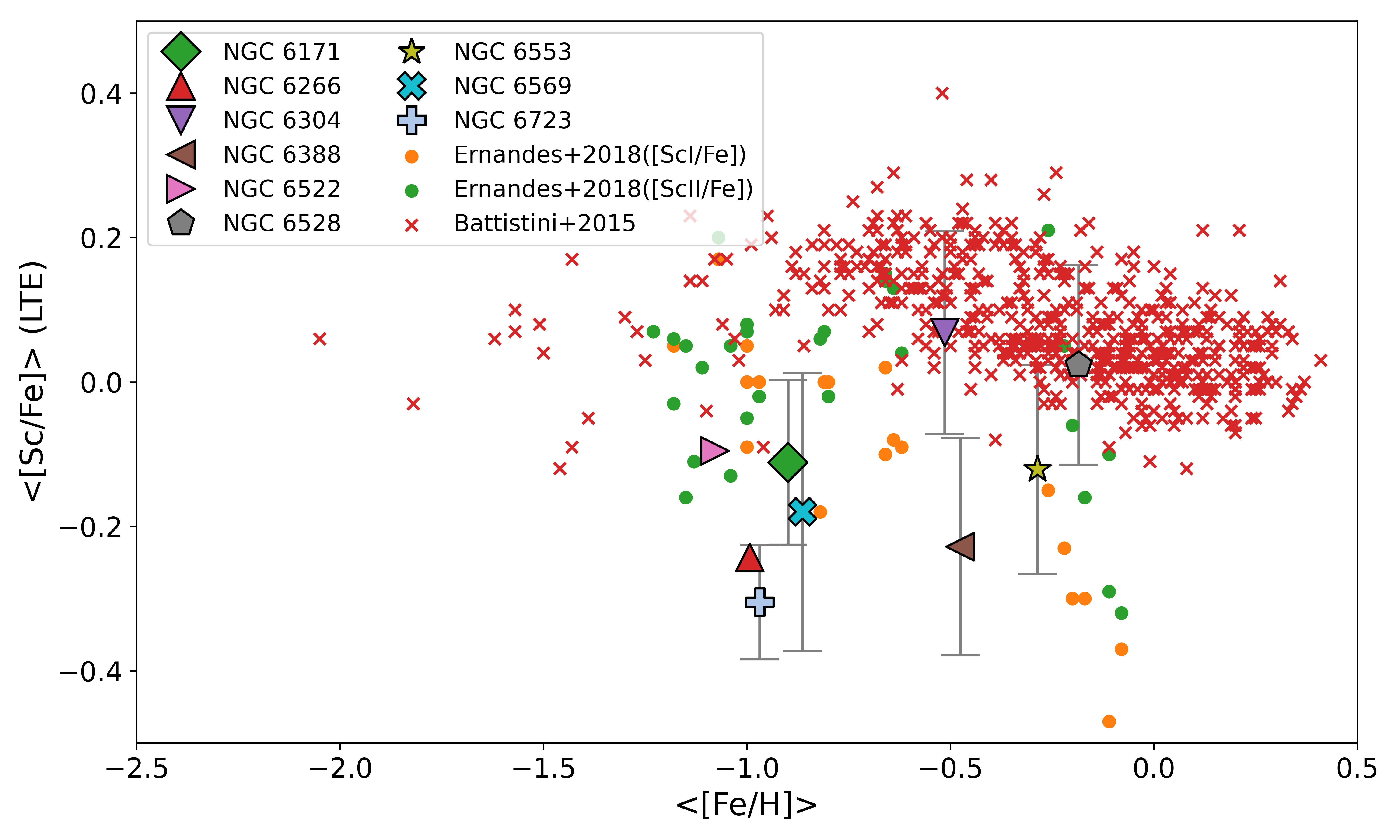}
    \caption{Mean [Sc/Fe] abundances, with their standard deviation, as a function of the mean metallicity of the clusters. The Sc abundances of thin and thick disk stars from \cite{Battistini2015} and of bulge cluster stars from \citet{Ernandes2018} are shown for comparison.}
    \label{fig:Sc_comp}
\end{figure}

Scandium is between typical $\alpha$- and iron-peak elements and is thought to be mainly produced by massive stars. Questions about its nucleosynthesis remain and it is known to be underproduced by theoretical models with respect to observations \citep[][and references therein]{Kobayashi2020}. Figure \ref{fig:Sc_comp} shows the mean cluster abundances we derived compared to the abundances from  \cite{Ernandes2018} and \cite{Battistini2015}, which studied Sc abundances in bulge globular clusters and disk stars, respectively. We only used \ion{Sc}{i} lines, while \cite{Ernandes2018} reported their derived abundances from both neutral and ionized Sc lines. Our clusters show abundances below the solar value and slightly lower than what \cite{Ernandes2018} reported for their sample. The trend we observe is very different from that seen in the sample of disk stars. This points to a different Sc contribution with respect to disk stars of the same metallicity; this behavior could potentially be used in models to help constrain the nucleosynthesis details of Sc.

\subsection{Heavy element abundances}

We derived abundances for Y, Zr, Ba, Ce, Nd, and Eu. For such heavy elements, the majority of GCs have similar abundances to halo stars. The exceptions are the so-called Type II GCs that display dispersion in metallicities and heavy elements \citep[see][for the classification details]{Milone2017}.

Elements heavier than iron are mostly produced by neutron-capture processes that can be classified according to properties of the environment in which they occur. In low-neutron-density environments, the nuclei capturing a neutron have enough time to undergo beta-decay before they encounter another neutron. This so-called slow process (s-process) creates stable elements sequentially along a stability valley, producing elements such as Sr and Ba. Low-mass AGB stars and low metallicity massive stars are the main production sites for s-process elements \citep[see review by][]{Lugaro2023}. The rapid neutron-capture process (r-process), on the other hand, takes place in high-neutron density environments where the time-scale for capturing neutrons is shorter than the beta decay. The r-process produces heavier elements and heavier isotopes of elements produced by the s-process (Eu is a typical tracer of the r-process). Supernovae/hypernovae and neutron star mergers are believed to be the main sites of the r-process \citep[see review by][]{Cowan2021}.

\begin{figure}
    \centering
    \includegraphics[width=\linewidth]{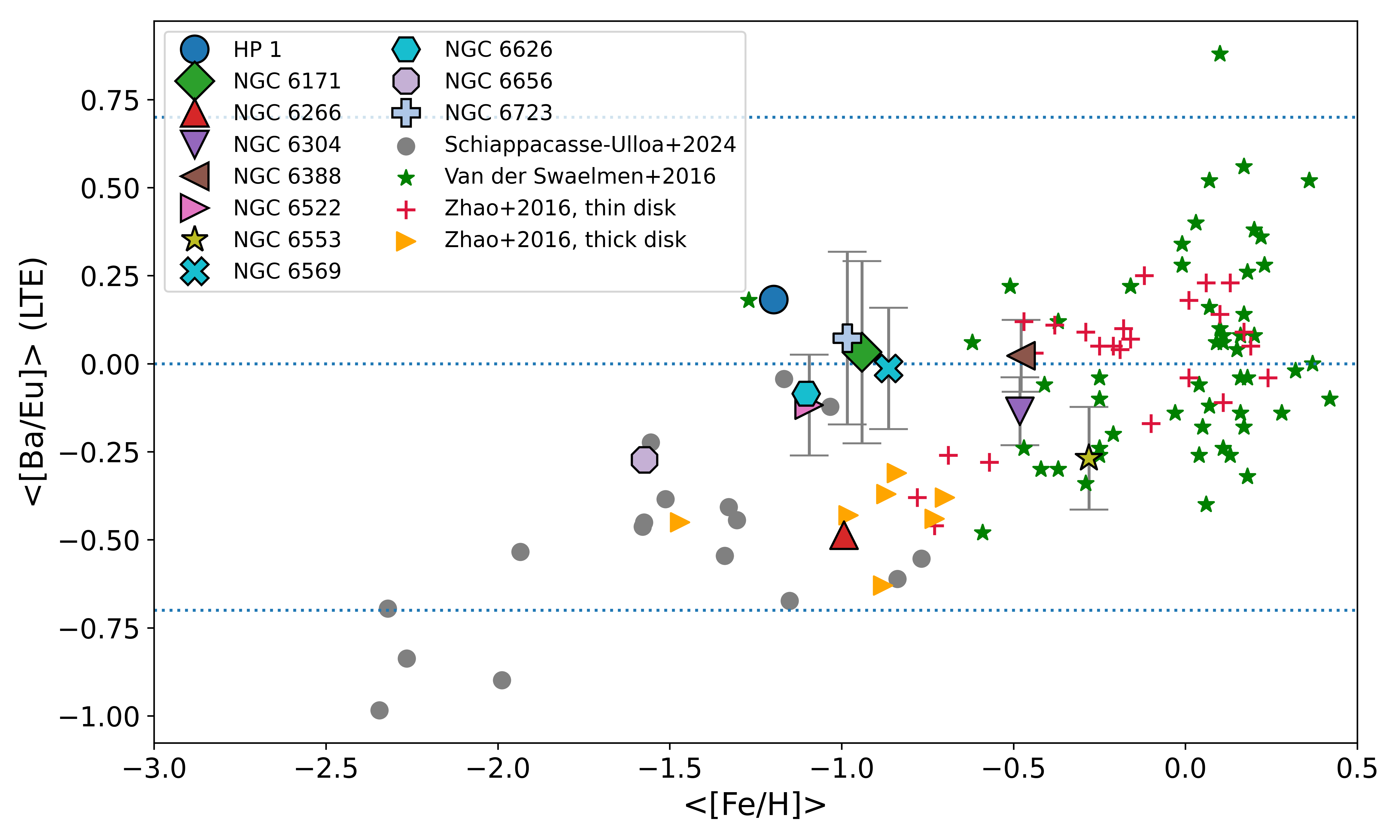}
    \caption{[Ba/Eu] vs. [Fe/H] for our sample in comparison to halo GCs from \cite{Schiappacasse24}, bulge stars from \cite{VanderSwaelmen2016}, and thin and thick disk stars from \citet{Zhao2016}. The vertical dashed lines at $-$0.7 and 0.7 show the levels for enrichment dominated by the r-process and s-process, respectively. The line at 0.0 shows the solar ratio.}
    \label{fig:BaEu}
\end{figure}

Massive stars evolve much faster than the time it takes for low-mass stars to reach the AGB. Therefore, a significant contribution from the r-process occurs before a significant addition of elements produced by the s-process. High abundances of r-process elements seen together with low abundances of s-process elements indicate a rapid enrichment at early times. In Figure \ref{fig:BaEu}, we plot our mean [Ba/Eu] abundances as a function of metallicity. Barium is an element typically produced by the s-process while Eu is a typical product of the r-process. We compared our ratios with mean GC values from \cite{Schiappacasse24}, bulge field stars from \cite{VanderSwaelmen2016}, and disk stars from \citet{Zhao2016}. As can be seen, bulge clusters show values around solar values that are higher than typically seen in low metallicity halo GCs and in thick disk stars. Instead, our abundances seem to follow the same trend seen in bulge field giants towards low metallicities. This implies that bulge clusters, similar to bulge field stars, have a significant contribution from s-process elements, either from AGB stars or low-metallicity massive stars. Instead, the halo clusters formed quickly from an environment enriched mainly by the r-process.

\begin{figure}
    \centering
    \includegraphics[width=\linewidth]{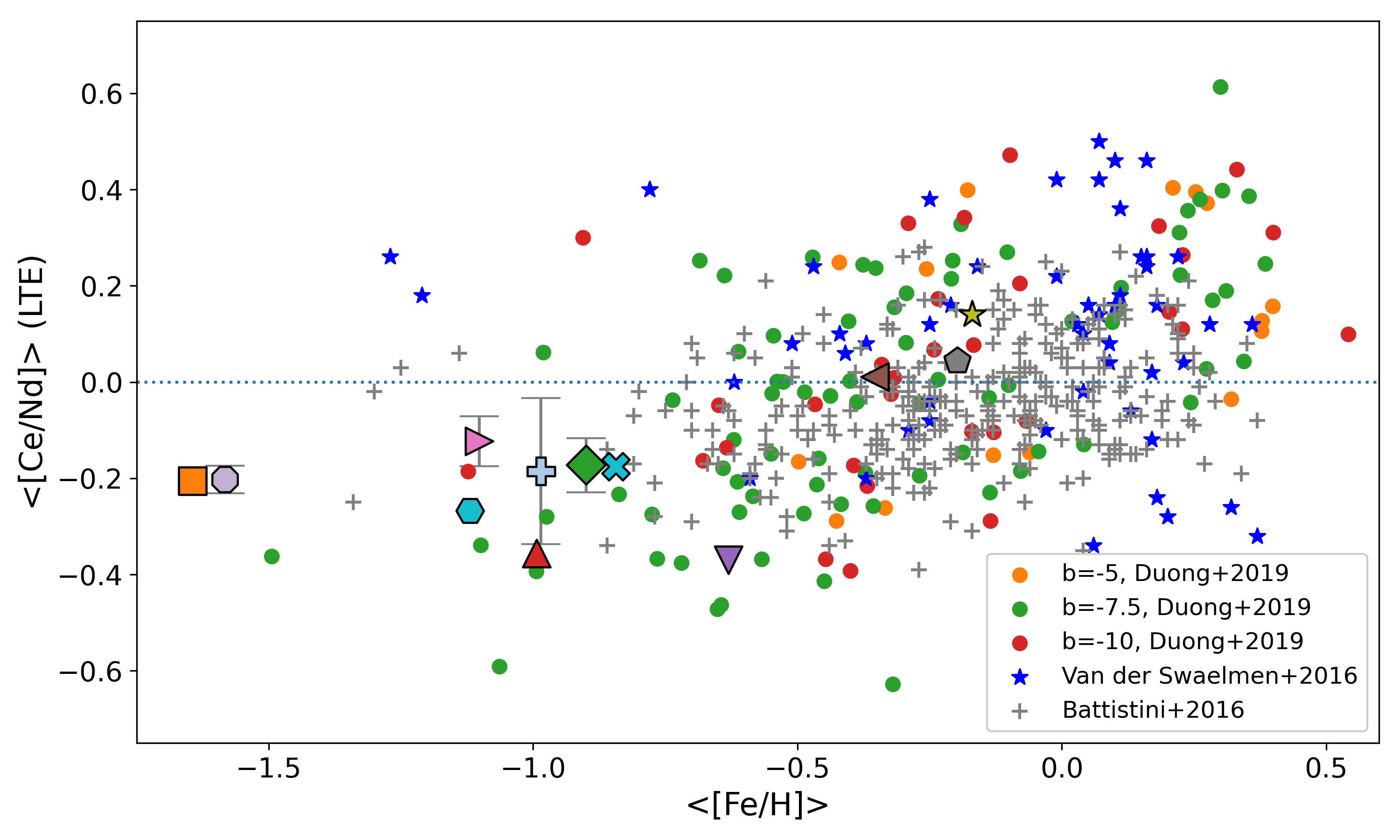}
    \caption{[Ce/Nd] as a function of metallicity for our sample in comparison to bulge giants \citep[][``b'' in the legend represents the Galactic latitude of the field]{VanderSwaelmen2016, Duong2019} and main sequence disk stars \citep{Battistini2016}.}
    \label{fig:CeNd}
\end{figure}

\begin{figure}
    \centering
    \includegraphics[width=\linewidth]{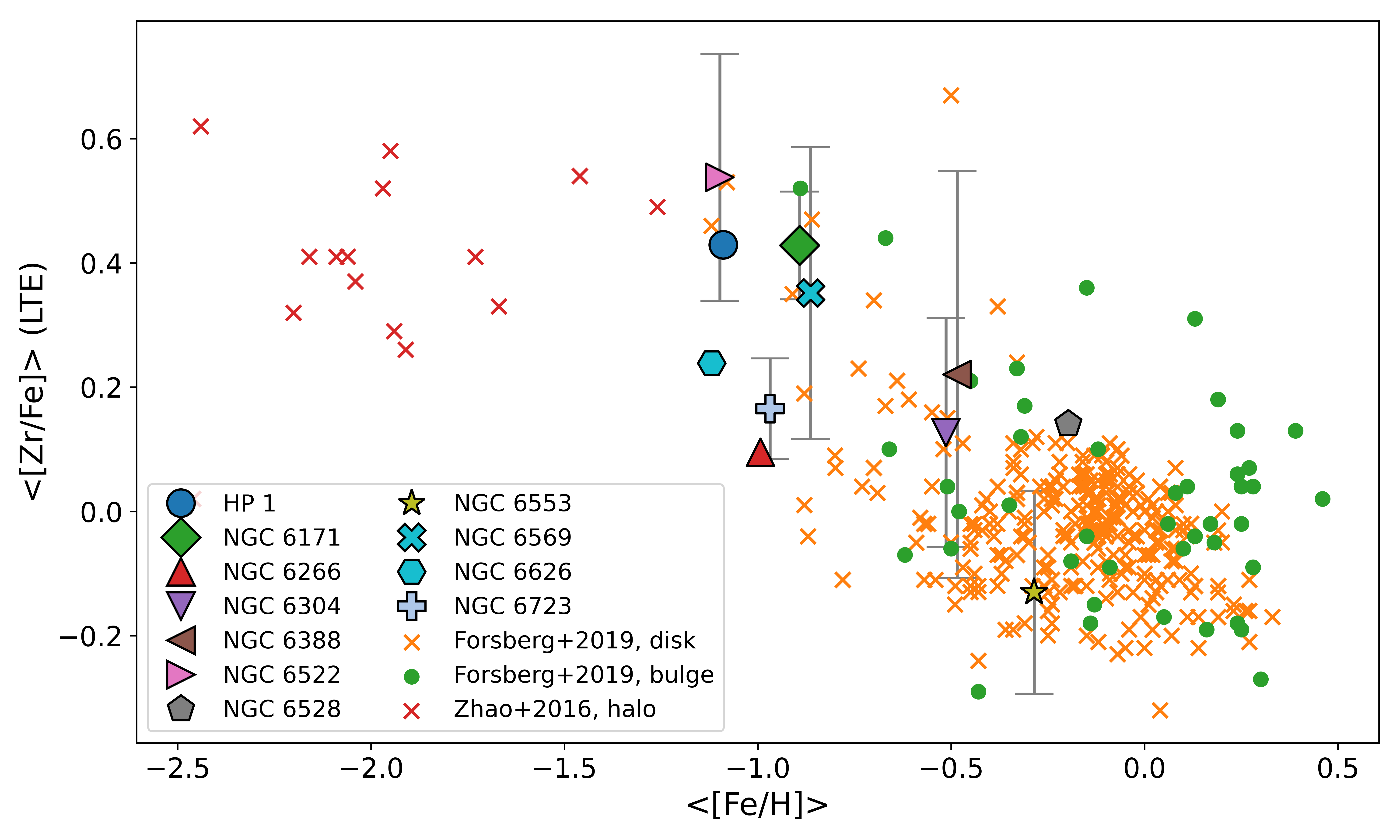}
    \caption{[Zr/Fe] as a function of metallicity for our sample for halo main sequence stars from \cite{Zhao2016} and for bulge and disk giants from \cite{Forsberg2019}.}
    \label{fig:Zr}
\end{figure}

In addition to Ba and Eu, the elements Ce and Nd have different enrichment contributions from neutron-capture processes \citep{Prantzos2020, SalesSilva2024}. The [Ce/Nd] ratio can also be used as a sensitive tool of the dominant process that contributes to heavy-element abundances in stars. Both the s- and r-process can produce Zr, Ce, and Nd \citep{Grisoni2020} but at different fractions; at solar metallicity, Zr, Ce, and Nd are produced at a level of 82\%, 85\%, and 62\% by the s-process, respectively \citep{Prantzos2020}. Therefore, a high value for [Ce/Nd] indicates a stronger contribution from the s-process, while a low ratio indicates dominance of the r-process \citep{SalesSilva2024}. As can be seen in Fig.~\ref{fig:CeNd}, our values of [Ce/Nd] show a trend that is similar to the one observed by \cite{Duong2019} in bulge giants, an increasing [Ce/Nd] ratio with increasing metallicity. This is unlike what was observed by \cite{VanderSwaelmen2016}. For bulge GC stars, this supports the claim from \cite{SalesSilva2024} that the r-process dominated the yields of Ce and Nd. This conclusion seems to apply to the metal-poor bulge globular clusters in our sample.

The observed Ce-Nd and Ba-Eu abundance trends might be seen in a conflict. The Ba abundances derived in this work are consistent with the stochastic chemical evolution of the bulge \citep{Cescutti2018}, which was derived as an extreme case of the halo model from \cite{Cescutti2014}. We also note that Eu might not be a robust indicator of the r-process in all formation sites, considering the possible decoupling observed between Eu and abundances of third-peak r-process elements seen by \cite{Puls2025}.

Finally, Figure \ref{fig:Zr} shows the comparison between the mean [Zr/Fe] values for each cluster and the sample of disk and bulge stars from \cite{Forsberg2019}. It is clear that both disk and bulge giants share a similar pattern for Zr, a decreasing trend of [Zr/Fe] values with increasing metallicity. In general, our sample follows the same pattern. This confirms that s-process elements alone do not seem to be able to disentangle thick disk, thin disk and bulge populations \citep[see also][]{Grisoni2020}. Studies with a higher level of precision might be needed to unravel differences in s-process abundance patterns between Galactic components.

\section{Summary and conclusions}\label{sec:summary}

We used a new pipeline, CHESS \citep{CHESSI}, for a homogeneous (re-)analysis of high resolution UVES archival spectra of giants in bulge GCs. We derived atmospheric parameters on an accurate scale for 135 stars from 14 bulge GCs and chemical abundances for 114 stars from 13 GCs (restricting the initial sample to S/N $>20$). This work is part of a series of papers in which we plan to discuss the abundances of FGK-type stars from different Galactic components, based on a homogeneous analysis of high-resolution spectra using methods designed to ensure both the accuracy and precision of the results \citep[][Smiljanic et al., in prep.]{CHESSI}. Our results for the bulge GCs we studied are summarized below.

\begin{itemize}
    \item Despite the low S/N of most of our data (61\% with 20 $\leq$ S/N $\leq$ 40), we were able to derive atmospheric parameters with errors that are typically better than 66 K, 0.22 dex, 0.08 dex, and 0.14 km s$^{-1}$ for \teff\, \logg\, [Fe/H], and \vmicro, respectively. Moreover, the parameter scale is anchored on a set of reference stars with accurate parameters.
    \item We derived abundances for 14 elements (Li, O, Na, Mg, Al, Si, Ca, Sc, Y, Zr, Ba, Ce, Nd, and Eu) using a spectral synthesis study, with different levels of completeness per star. We took into account the NLTE effects for Na, Mg, Al, Si, Ca, Y, Ba, and Eu.
    \item We observed a Na spread that confirms the presence of multiple populations in NGC 6171, NGC 6304, NGC 6388, NGC 6522, NGC 6553, and NGC 6723. Their spread in Na is larger than the errors in the analysis, indicating that it is not caused by the low S/N and that it is a real sign of different populations. Our limited investigation of Mg and Al indicates a possible anti-correlation in NGC 6522. 
    \item We discovered a Li-rich giant star in NGC 6723, the 16th known in GCs. Its parameters and abundances suggest that it is a first-generation star in which the Li enrichment is as a result of intrinsic stellar evolution. A more detailed study of this object would be useful in supporting that interpretation. This discovery was made by exploiting archival data, highlighting the potential that may be hidden in data archives;
    \item For low metallicities, the [$\alpha$/Fe] plateau in our bulge GCs is below what is seen in halo clusters. The ratio then decreases with increasing metallicity (starting around [Fe/H]$\sim -$0.7 dex). This decrease is milder than what is typically seen for halo clusters. In other words, at higher metallicities, the [$\alpha$/Fe] of bulge GCs remains higher than in halo clusters.
    \item The [Sc/Fe] ratio in bulge GCs is lower than that seen in disk stars of the same metallicity. This difference could be useful for constraining models of the nucleosynthesis of Sc.
     \item We do not see a significant spread in heavy elements in our sample. NGC 6656 (M22, which is a likely halo intruder) is the only cluster with a significant spread in Y and Ba, as previously known, showing the complex nature of this cluster. We note a tentative Ba spread for HP 1, but the small number of stars (three) makes it difficult to reach a firm conclusion.
    \item Our abundances support the \cite{SalesSilva2024} conclusion that in metal-poor bulge GCs the elements Ce and Nd show a significant contribution from the r-process. A dedicated study on this issue would be important as means to better constrain chemical evolution models.
    \item The bulge GCs show near-solar [Ba/Eu] ratios, indicating that we are detecting enrichment from both the r- and s-processes. This is consistent with what is seen in bulge field stars but different from very metal-poor halo GCs. In principle, this might be seen to conflict with the conclusions drawn for Ce and Nd. The observed Ba abundances in this study are in agreement with chemical evolution models of the bulge, but we note that Eu might not be a robust indicator for r-process in every formation site.
\end{itemize}

Our homogeneous analysis of high-resolution archival spectra of stars belonging to Galactic bulge GCs provides key information for models of bulge chemical evolution. The serendipitous discovery of a Li-rich giant shows the importance of systematically exploiting existing spectroscopic archives.

\section*{Data Availability}
Table E.1 is only available in electronic form at the CDS via anonymous ftp to \url{cdsarc.u-strasbg.fr} (130.79.128.5) or via \url{http://cdsweb.u-strasbg.fr/cgi-bin/qcat?J/A+A/}.

\begin{acknowledgements}
S. \"Ozdemir, J.~E.~Mart\'{i}nez Fern\'andez, and R. Smiljanic acknowledge support from the National Science Centre, Poland, project 2019/34/E/ST9/00133. PC acknowledgments the support by Conselho Nacional de Desenvolvimento Científico e Tecnológico (Process 308715/2025-0). Based on data obtained from the ESO Science Archive Facility with DOI: \url{https://doi.eso.org/10.18727/archive/50} \citep{UVESreduceddata}. Based on observations collected at the European Southern Observatory under ESO programme(s) Mention Programme ids of ESO: 067.B-0382, 068.D-0332, 071.B-0617, 073.D-0211, 073.D-0760, 075.D-0043, 077.B-0327, 087.D-0230, 087.D-0344, 088.D-0398, 091.D-0151, 091.D-0383, 091.D-0535, 093.D-0124, 093.D-0286, 095.D-0834, 097.D-0175, 097.D-1017, 099.D-0047, 099.D-0263, 101.D-0620, 193.D-0232. This research has made use of the SIMBAD database, operated at CDS, Strasbourg, France. This work has made use of data from the European Space Agency (ESA) mission {\it Gaia} (\url{https://www.cosmos.esa.int/gaia}), processed by the {\it Gaia} Data Processing and Analysis Consortium (DPAC, \url{https://www.cosmos.esa.int/web/gaia/dpac/consortium}). Funding for the DPAC has been provided by national institutions, in particular the institutions participating in the {\it Gaia} Multilateral Agreement.
\end{acknowledgements}

\bibliographystyle{aa}
\bibliography{aa62623-26.bib}

\begin{appendix}

\section{Functions for bias correction and the relation for microturbulence velocity.}\label{ap:bias}

The equations to calculate the bias that should be applied to the atmospheric parameters to put them on an accurate scale defined by the reference stars are summarized in Table ~\ref{apx:tab:bias_corrections}. They assume as input the normalized parameters computed from Equations ~\ref{apx:eq:teff_norm}, ~\ref{apx:eq:logg_norm}, and  ~\ref{apx:eq:metal_norm}, expressed as

\begin{equation}\label{apx:eq:teff_norm}
    T_{\textrm{eff, norm}} = (T_{\textrm{eff}} - 5178) / 563, 
\end{equation}

\begin{equation}\label{apx:eq:logg_norm}
    \logg_{\textrm{norm}} = (\logg - 2.92) / 0.77, 
\end{equation}

\begin{equation}\label{apx:eq:metal_norm}
    \rm{[Fe/H]}_{\textrm{norm}} = ( \rm{[Fe/H]} + 0.68) / 0.94. 
\end{equation}

\begin{table}
\caption{Equations to calculate the bias correction values to be applied on \teff, \logg, and [Fe/H].}
\label{apx:tab:bias_corrections}
$$
\begin{array}{lrrr}
\hline \hline
\rm{Term} & \rm{Bias}\ {\it T}_{\rm{eff}} & \rm{Bias}\ \logg & \rm{Bias}\ \rm{[Fe/H]} \\
\hline
\rm{Intercept} & 54.08 & 0.107 & 0.033 \\
T_{\textrm{eff, norm}} & -61.51 & -0.018 & 0.021 \\
T_{\textrm{eff, norm}}^2 & - & 0.044 & 0.041 \\
\logg_{\textrm{norm}} & 101.51 & 0.132 & - \\
\logg_{\textrm{norm}}^2 & - & 0.053 & 0.063 \\
\rm{[Fe/H]}_{\textrm{norm}} & -158.51 & -0.280 & -0.096 \\
\rm{[Fe/H]}_{\textrm{norm}}^2 & 25.00 & 0.049 & 0.053 \\
T_{\textrm{eff, norm}} \times \logg_{\textrm{norm}} & - & -0.060 & -0.091 \\
T_{\textrm{eff, norm}} \times \rm{[Fe/H]}_{\textrm{norm}} & -33.61 & -0.132 & 0.010 \\
\logg_{\textrm{norm}} \times \rm{[Fe/H]}_{\textrm{norm}}& 10.18 & 0.076 & -0.052 \\
\hline
\end{array}
$$
\end{table}

Microturbulence velocities are calculated with the empirical relation in Equation \ref{apx:eq:vt_relation} using the bias-corrected parameters. This function was derived using the \textit{Gaia}-ESO results for giant stars (\logg $<$ 3.5 dex). The complete set of relations for other stellar types and a discussion of their derivation will be given in Smiljanic et al. (in preparation).

\begin{equation}
\label{apx:eq:vt_relation}
\begin{split}
\vmicro = 1.47 & + 4.58 \times 10^{-4} \Delta T_{\mathrm{eff}} + 2.16 \times 10^{-7} \Delta T_{\mathrm{eff}}^2 \\
           & - 5.08 \times 10^{-1} \Delta \logg - 7.71 \times 10^{-2} \Delta \logg^2 \\
           & + 2.20 \times 10^{-1} \Delta \mathrm{[Fe/H]} + 5.45 \times 10^{-2} \Delta \mathrm{[Fe/H]}^2
\end{split}
\end{equation}

\noindent where the atmospheric parameters are defined relative to reference offsets:
\begin{align*}
\Delta T_{\mathrm{eff}} &= T_{\mathrm{eff}} - 4798 \\
\Delta \log g           &= \log g - 2.38 \\
\Delta \mathrm{[Fe/H]}  &= \mathrm{[Fe/H]} + 0.71
\end{align*}

\section{Linelist used for chemical abundances} \label{apx:elemental_lines}

In Table \ref{tab:apx:linelist}, we list the lines that were used for abundance analysis in this work. Atomic data from \textit{Gaia}-ESO \citep{Heiter21} were used, except for Mg, Si, Ba, and Eu. For the Ba and Eu lines we used the hyperfine structure and isotopic shift from \cite{daSilva_Smiljanic_2025}. The updated atomic data for the Mg and Si lines of \cite{Magg2022} were used, as implemented in the default TSFitPy line list. 

\begin{table}
\centering
\caption{Lines that were used for elemental abundances}
\label{tab:apx:linelist}
\begin{tabular}{ccc}
\hline \hline
Species &N& $\lambda$  \\
        & &   \AA  \\
\hline
 Li I     & 1 & 6707.8 \\
 {[}O I] & 1 & 6300.3 \\
 Na I     & 2 & 6154.2; 6160.7 \\
 Mg I     & 2 & 5528.4; 5711.1 \\
 Al I     & 2 & 6696.0; 6698.7 \\
 Si I     & 4 & 5690.4; 5701.1; 5948.5; 6721.8\\
 Ca I     & 8 & 5260.4; 5588.7; 5590.1; 6122.2 \\
          &   & 6162.2; 6166.4; 6169.0; 6169.6\\
 Sc I     & 4 & 5484.6; 5520.5; 5686.8; 6210.7\\
 Y II     & 2 & 5087.4; 5200.4 \\
 Zr I     & 3 & 6127.4; 6134.6; 6143.2 \\
 Ba II    & 3 & 5853.7; 6141.7; 6496.9 \\
 Ce II    & 1 & 5274.2 \\
 Nd II    & 3 & 5089.8; 5092.8; 5249.6\\
 Eu II    & 1 & 6645.1 \\
\hline
\end{tabular}
\end{table}

\section{Elemental abundances of bulge clusters}\label{apx:elements}

In Table ~\ref{apx:tab:light_elements}, we summarize the mean abundances of light elements in LTE for each cluster. Similarly, Table ~\ref{apx:tab:heavy_elements} summarizes the mean abundances for the heavy elements, also in LTE. Abundances in NLTE are given in Tables \ref{apx:tab:light_elements_nlte} and \ref{apx:tab:heavy_elements_nlte} for the light and heavy elements, respectively.

\begin{table*}
\caption{Mean LTE light element abundances and their standard deviations for each cluster. Numbers inside the parentheses denote the number of stars used for the given abundance.}
\label{apx:tab:light_elements}
$$
\begin{array}{ccccccc}
\hline \hline
\rm{Cluster} & \rm{<[Na/Fe]>} & \rm{<[Mg/Fe]>} & \rm{<[Al/Fe]>} & \rm{<[Si/Fe]>} & \rm{<[Ca/Fe]>} & \rm{<[Sc/Fe]>} \\
\hline
\rm{HP\ 1} & 0.53\ (1) & - & - & - & 0.22 \pm 0.02(3) & - \\
\rm{NGC\ 6093} & - & 0.34 \pm 0.04(2) & - & 0.32 \pm 0.09(2) & 0.39 \pm 0.13(7) & - \\
\rm{NGC\ 6171} & 0.24 \pm 0.27(15) & 0.47 \pm 0.12(16) & 0.49 \pm 0.09(8) & 0.39 \pm 0.09(16) & 0.24 \pm 0.09(16) & -0.11 \pm 0.11(7) \\
\rm{NGC\ 6266} & 0.05\ (1) & 0.25 \pm 0.05(2) & 0.12\ (1) & 0.22\ (1) & 0.02 \pm 0.02(2) & -0.24\ (1) \\
\rm{NGC\ 6304} & 0.58 \pm 0.24(11) & 0.44 \pm 0.10(7) & 0.50 \pm 0.10(10) & 0.35 \pm 0.14(11) & 0.15 \pm 0.11(11) & 0.07 \pm 0.14(11) \\
\rm{NGC\ 6388} & 0.33 \pm 0.26(8) & 0.14 \pm 0.04(5) & 0.47 \pm 0.28(7) & 0.20 \pm 0.13(6) & 0.10 \pm 0.19(9) & -0.23 \pm 0.15(7) \\
\rm{NGC\ 6522} & 0.21 \pm 0.31(10) & 0.35 \pm 0.07(10) & 0.61 \pm 0.25(5) & 0.27 \pm 0.08(12) & 0.20 \pm 0.09(12) & -0.10\ (1) \\
\rm{NGC\ 6528} & 0.53 \pm 0.19(3) & 0.23\ (1) & 0.52 \pm 0.19(2) & 0.47 \pm 0.11(2) & 0.18 \pm 0.12(4) & 0.02 \pm 0.14(4) \\
\rm{NGC\ 6553} & 0.37 \pm 0.23(8) & 0.26 \pm 0.12(8) & 0.41 \pm 0.09(8) & 0.35 \pm 0.09(6) & 0.00 \pm 0.10(8) & -0.12 \pm 0.14(8) \\
\rm{NGC\ 6569} & 0.06 \pm 0.16(4) & 0.31 \pm 0.07(4) & 0.38 \pm 0.06(3) & 0.25 \pm 0.04(4) & 0.13 \pm 0.13(5) & -0.18 \pm 0.19(2) \\
\rm{NGC\ 6626} & 0.46 \pm 0.01(2) & 0.36 \pm 0.05(4) & - & 0.29 \pm 0.09(4) & 0.21 \pm 0.06(4) & - \\
\rm{NGC\ 6656} & - & 0.34 \pm 0.04(4) & - & 0.33 \pm 0.15(4) & 0.21 \pm 0.04(4) & - \\
\rm{NGC\ 6723} & 0.03 \pm 0.15(22) & 0.40 \pm 0.07(23) & 0.19 \pm 0.09(3) & 0.25 \pm 0.06(22) & 0.14 \pm 0.05(23) & -0.30 \pm 0.08(4) \\
\hline
\end{array}
$$
\end{table*}

\begin{table*}
\caption{The derived mean LTE heavy element abundances in the same formatting as Table ~\ref{apx:tab:light_elements}.}
\label{apx:tab:heavy_elements}
$$
\begin{array}{ccccccc}
\hline \hline
\rm{Cluster} & \rm{<[Y/Fe]>} & \rm{<[Zr/Fe]>} & \rm{<[Ba/Fe]>} & \rm{<[Ce/Fe]>} & \rm{<[Nd/Fe]>} & \rm{<[Eu/Fe]>} \\
\hline
\rm{HP\ 1} & - & 0.43\ (1) & 0.47 \pm 0.30(2) & - & - & 0.50\ (1)  \\
\rm{NGC\ 6093} & -0.17 \pm 0.13(3) & - & 0.46 \pm 0.21(2) & 0.01\ (1) & 0.44 \pm 0.15(4) & -  \\
\rm{NGC\ 6171} & 0.07 \pm 0.19(13) & 0.43 \pm 0.09(6) & 0.59 \pm 0.17(16) & 0.12 \pm 0.09(7) & 0.32 \pm 0.08(10) & 0.56 \pm 0.14(11)  \\
\rm{NGC\ 6266} & -0.40 \pm 0.08(2) & 0.09\ (1) & -0.14 \pm 0.14(2) & -0.37\ (1) & -0.08 \pm 0.10(2) & 0.45\ (1)  \\
\rm{NGC\ 6304} & -0.44\ (1) & 0.13 \pm 0.18(11) & 0.31 \pm 0.12(9) & -0.28 \pm 0.11(6) & 0.22\ (1) & 0.41 \pm 0.11(8) \\
\rm{NGC\ 6388} & - & 0.22 \pm 0.33(8) & 0.42 \pm 0.23(5) & -0.23 \pm 0.08(5) & -0.33\ (1) & 0.29 \pm 0.10(7) \\
\rm{NGC\ 6522} & -0.13 \pm 0.19(6) & 0.54 \pm 0.20(3) & 0.25 \pm 0.15(11) & -0.07 \pm 0.06(2) & 0.04 \pm 0.04(3) & 0.36 \pm 0.06(6) \\
\rm{NGC\ 6528} & - & 0.14\ (1) & 0.34 \pm 0.04(3) & -0.07\ (1) & -0.11\ (1) & - \\
\rm{NGC\ 6553} & - & -0.13 \pm 0.16(8) & 0.07 \pm 0.14(3) & -0.15 \pm 0.08(4) & -0.32 \pm 0.02(2) & 0.25 \pm 0.10(7)  \\
\rm{NGC\ 6569} & 0.20\ (1) & 0.35 \pm 0.23(2) & 0.51 \pm 0.17(4) & -0.15\ (1) & 0.10 \pm 0.11(2) & 0.49 \pm 0.10(2)  \\
\rm{NGC\ 6626} & -0.17 \pm 0.13(4) & 0.24\ (1) & 0.34 \pm 0.12(4) & -0.18\ (1) & 0.12 \pm 0.05(2) & 0.41\ (1) \\
\rm{NGC\ 6656} & -0.18 \pm 0.29(4) & - & 0.39 \pm 0.42(4) & -0.02 \pm 0.16(3) & 0.17 \pm 0.13(4) & 0.32\ (1) \\
\rm{NGC\ 6723} & -0.10 \pm 0.13(22) & 0.17 \pm 0.08(4) & 0.42 \pm 0.20(23) & -0.22 \pm 0.16(8) & 0.01 \pm 0.13(16) & 0.30 \pm 0.12(10) \\
\hline
\end{array}
$$
\end{table*}

\begin{table*}
\caption{Same as Table \ref{apx:tab:light_elements} for NLTE abundances.}
\label{apx:tab:light_elements_nlte}
$$
\begin{array}{cccccc}
\hline \hline
\rm{Cluster} & \rm{<[Na/Fe]>} & \rm{<[Mg/Fe]>} & \rm{<[Al/Fe]>} & \rm{<[Si/Fe]>} & \rm{<[Ca/Fe]>}  \\
\hline
\rm{HP\ 1}	   &  0.50 (1)   &	-	 & -    & -    & 0.20 \pm 0.01(3)  \\
\rm{NGC\ 6093} &	-	&  0.15 \pm 0.05(2)	 & -    & 0.11 \pm 0.08(2)  & 0.35 \pm 0.14(7)  \\
\rm{NGC\ 6171} & 0.21 \pm 0.27(15)  &  0.29 \pm 0.11(16) & 0.43 \pm 0.09(8)  & 0.25 \pm 0.08(16) & 0.23 \pm 0.09(16) \\
\rm{NGC\ 6266} & 0.04 (1)    &  0.06 \pm 0.02(2)  & 0.06 (1)   & 0.10 (1)   & 0.02 \pm 0.02(2)  \\
\rm{NGC\ 6304} & 0.47 \pm 0.21(11)  &  0.31 \pm 0.10(7)  & 0.42 \pm 0.10(10) & 0.22 \pm 0.13(11) & 0.14 \pm 0.11(11) \\
\rm{NGC\ 6388} & 0.24 \pm 0.22(8)   &  0.01 \pm 0.03(5)  & 0.39 \pm 0.27(7)  & 0.10 \pm 0.13(6)  & 0.10 \pm 0.18(9)  \\
\rm{NGC\ 6522} & 0.17 \pm 0.32(10)  &  0.16 \pm 0.07(10) & 0.56 \pm 0.25(5)  & 0.12 \pm 0.09(12) & 0.19 \pm 0.09(12) \\
\rm{NGC\ 6528} & 0.43 \pm 0.17(3)   &  0.14 (1)   & 0.44 \pm 0.14(2)  & 0.37 \pm 0.07(2)  & 0.19 \pm 0.12(4)  \\
\rm{NGC\ 6553} & 0.27 \pm 0.20(8)   &  0.15 \pm 0.11(8)  & 0.32 \pm 0.07(8)  & 0.29 \pm 0.09(6)  & 0.01 \pm 0.10(8)  \\
\rm{NGC\ 6569} & 0.05 \pm 0.15(4)   &  0.15 \pm 0.06(4)  & 0.32 \pm 0.08(3)  & 0.15 \pm 0.03(4)  & 0.12 \pm 0.13(5)  \\
\rm{NGC\ 6626} & 0.44 \pm 0.01(2)   &  0.15 \pm 0.04(4)  & -    & 0.13 \pm 0.09(4)  & 0.18 \pm 0.06(4)  \\
\rm{NGC\ 6656} & -	    &  0.14 \pm 0.04(4)  & -    & 0.14 \pm 0.12(4)  & 0.19 \pm 0.04(4)  \\
\rm{NGC\ 6723} & 0.01 \pm 0.15(22)  &  0.18 \pm 0.06(23) & 0.13 \pm 0.09(3)  & 0.10 \pm 0.05(22) & 0.13 \pm 0.05(23) \\
\hline
\end{array}
$$
\end{table*}

\begin{table*}
\caption{The derived mean NLTE heavy element abundances in the same formatting as Table ~\ref{apx:tab:light_elements}.}
\label{apx:tab:heavy_elements_nlte}
$$
\begin{array}{cccc}
\hline \hline
\rm{Cluster} & \rm{<[Y/Fe]>} & \rm{<[Ba/Fe]>} & \rm{<[Eu/Fe]>} \\
\hline
\rm{HP\ 1}     &           -        & 0.26 \pm 0.21(2)  &     0.36 (1)      \\
\rm{NGC\ 6093} & -0.10 \pm 0.12(3)  & 0.22 \pm 0.24(2)  &      -            \\
\rm{NGC\ 6171} & 0.12 \pm 0.15(13)  & 0.45 \pm 0.19(16) & 0.48 \pm 0.15(11) \\
\rm{NGC\ 6266} & -0.32 \pm 0.05(2)  & -0.30 \pm 0.25(2) &     0.38 (1)      \\
\rm{NGC\ 6304} &      -0.35 (1)     & 0.26 \pm 0.15(9)  & 0.35 \pm 0.10(8)  \\
\rm{NGC\ 6388} &          -         & 0.39 \pm 0.31(5)  & 0.25 \pm 0.10(7)  \\
\rm{NGC\ 6522} & -0.07 \pm 0.18(6)  & 0.05 \pm 0.16(11) & 0.25 \pm 0.07(6)  \\ 
\rm{NGC\ 6528} &          -         & 0.29 \pm 0.07(3)  &      -            \\
\rm{NGC\ 6553} &          -         & 0.05 \pm 0.15(3)  & 0.21 \pm 0.10(7)  \\
\rm{NGC\ 6569} &       0.25(1)      & 0.39 \pm 0.14(4)  & 0.44 \pm 0.08(2)  \\
\rm{NGC\ 6626} & -0.10 \pm 0.14(4)  & 0.09 \pm 0.15(4)  &     0.30 (1)      \\
\rm{NGC\ 6656} & -0.12 \pm 0.28(4)  & 0.15 \pm 0.38(4)  &     0.19 (1)      \\
\rm{NGC\ 6723} & -0.04 \pm 0.12(22) & 0.24 \pm 0.21(23) & 0.18 \pm 0.14(10) \\
\hline
\end{array}
$$
\end{table*}

\section{Abundance uncertainties}\label{apx:errors}

Table ~\ref{tab:apx:errors} summarizes the sensitivity of the derived abundances to changes in atmospheric parameters and the total error for each element. For this analysis, we adopted typical errors of 66 K, 0.22 dex, 0.08 dex and 0.14 kms$^{-1}$ for \teff\ , \logg\ , [Fe/H] and \vmicro\ , respectively.

\begin{table*}
\centering
\caption{Sensitivity and total errors of derived abundances.}
\label{tab:apx:errors}
\begin{tabular}{cccccc}
\hline \hline
\multicolumn{1}{c}{\multirow{2}{*}{Species}} & \multicolumn{1}{c}{$\Delta$\teff} & \multicolumn{1}{c}{$\Delta$\logg} & \multicolumn{1}{c}{$\Delta$[Fe/H]} & \multicolumn{1}{c}{$\Delta$\vmicro} & \multicolumn{1}{c}{Total error}\\
\multicolumn{1}{c}{} & $\pm$66 K & $\pm$0.22 dex & $\pm$0.08 dex & $\pm$0.14 kms$^{-1}$& $\sqrt{\sum x^2}$ \\
\hline
$\rm{[Na/Fe]}$ & 0.06 & 0.01 & 0.09 & 0.03 & 0.11 \\
$\rm{[Mg/Fe]}$ & 0.05 & 0.01 & 0.09 & 0.05 & 0.11 \\
$\rm{[Al/Fe]}$ & 0.05 & 0.00 & 0.08 & 0.02 & 0.10 \\
$\rm{[Si/Fe]}$ & 0.02 & 0.04 & 0.09 & 0.02 & 0.10 \\
$\rm{[Ca/Fe]}$ & 0.08 & 0.01 & 0.09 & 0.07 & 0.13 \\
$\rm{[Sc/Fe]}$ & 0.13 & 0.02 & 0.10 & 0.02 & 0.16 \\
$\rm{[Y/Fe]}$  & 0.02 & 0.09 & 0.07 & 0.06 & 0.12 \\
$\rm{[Zr/Fe]}$ & 0.13 & 0.03 & 0.09 & 0.03 & 0.16 \\
$\rm{[Ba/Fe]}$ & 0.05 & 0.08 & 0.05 & 0.13 & 0.16 \\
$\rm{[Ce/Fe]}$ & 0.01 & 0.10 & 0.07 & 0.01 & 0.12 \\
$\rm{[Nd/Fe]}$ & 0.02 & 0.10 & 0.07 & 0.01 & 0.12 \\
$\rm{[Eu/Fe]}$ & 0.01 & 0.10 & 0.07 & 0.01 & 0.12 \\
\hline
\end{tabular}
\end{table*}

\section{Stellar parameters and abundances}

Table E.1: Derived stellar parameters and elemental abundances for the whole sample with IDs (from CHESS target identification) and archival metadata. This table has been published online in CDS.

\end{appendix}

\end{document}